\pdfoutput=1

\documentclass[11pt,letterpaper]{article}
\usepackage{jheppub}
\usepackage{journals}
\usepackage{graphicx}
\usepackage{placeins}
\usepackage{multirow}
\usepackage[arrow, matrix, curve]{xy}

  \usepackage[english]{babel} 
  \usepackage{german}    
  \usepackage[utf8]{inputenc} 
  \usepackage{amsmath}   
  \usepackage{amsthm}    
  \usepackage{amsfonts}  
  \usepackage{relsize}   
  \usepackage{tikz}      
  \usepackage[mathscr]{euscript}
  \usepackage{amssymb}
  \usepackage[headinclude]{scrpage2}  
  \usepackage{graphicx}
  \usepackage{amsmath}
  \usepackage{slashed}
  \usepackage{lmodern}
   \usepackage{amsmath} 
   \usepackage{placeins}
  \usepackage{graphicx}
  \usepackage{wrapfig}
  \usepackage{subfig}
  \usepackage{pgfplots}
  \usepackage{amssymb}
  \usepackage[T1]{fontenc}
  \usepackage{setspace}  
  \usepackage{placeins}
\usepackage[arrow,curve,matrix]{xy}

  \newtheorem{satz}{Theorem}[section]
  \newtheorem{lemma}[satz]{Lemma}
  
  \theoremstyle{definition}
  
  \newtheorem{bemerkung}[satz]{comment}
  
  \newenvironment{beweis}
    {\begin{proof}[proof]}
    {\end{proof}}

  \makeatletter                    
    \def\blfootnote{\xdef\@thefnmark{}\@footnotetext}

\begin{document}

 \title		{Quasinormal Modes of magnetic black branes at finite 't Hooft coupling}

\author[a]	{Sebastian Waeber}
\affiliation[a] {Institute for Theoretical Physics, University of Regensburg,
		D-93040 Regensburg, Germany}
		\emailAdd	{sebastian.waeber@physik.uni-regensburg.de}
\abstract
    {The aim of this work is to extend the knowledge about Quasinormal Modes (QNMs) and the equilibration of strongly coupled systems, specifically of a quark gluon plasma (which we consider to be in a strong magnetic background field) by using the duality between $\mathcal{N}=4$ Super Yang-Mills (SYM) theory and type IIb Super Gravity (SUGRA) and including higher derivative corrections.   The  behaviour of the equilibrating system can be seen as the response of the system to tiny excitations. A quark gluon plasma in a strong magnetic background field, as produced for very short times during an actual heavy ion collision, is described holographically by certain metric solutions  to $5\text{D}$ Einstein-Maxwell-(Chern-Simons) theory, which can be obtained from  type IIb SUGRA. We are going to compute higher derivative corrections to this metric and consider $\alpha'^3$ corrections to tensor-quasinormal modes in this background geometry. We find indications for a strong influence of the magnetic background field on the equilibration behaviour also and especially when we include higher derivative corrections. }
\maketitle

\section{Introduction}
The formalism to qualitatively describe the early, far from equilibrium dynamics of the QCD phase of high energy density (for which  the term quark gluon plasma (QGP) will be used even in the non-thermalized state) generated during heavy ion collisions at RHIC or LHC is one of the most prominent applications of  gauge/gravity duality, more specifically of the duality between $\mathcal{N}=4$ super Yang-Mills theory (SYM) in $4$ dimensions and  supergravity (SUGRA) on  $\text{AdS}_5 \times S_5$ known as the AdS/CFT duality. In the weak limit of this duality the gauge group rank $N$ of the boundary theory is taken to infinity, while the 't Hooft coupling $\lambda$ is held fixed during the $N\to \infty$ limit and afterwards is taken to infinity as well.  This limit of the holographic duality allows for the description of far from equilibrium  dynamics at strong coupling. \\ \indent
After having determined certain observables within the AdS/CFT duality an interesting next question would be how their higher derivative or $\alpha'$-corrections  behave and how large they are. Computing finite 't Hooft coupling corrections  is notoriously messy and involved, but necessary, if ones wishes to leave the unrealistic $\lambda \to \infty$ limit.
\\ \indent Within a formalism that helps to describe QGPs far from equilibrium a natural aspect that should be analysed is how and how fast such a system equilibrates. At late times this question breaks down to the analysis of quasinormal modes (QNMs),  fluctuations around the equilibrium state. The inverse of the absolute value of the imaginary part of QNM frequencies, which correspond to the poles of the propagator of such a fluctuation, is proportional to the equilibration time. Thus, the QNM with the smallest absolute imaginary part determines the time the system needs to equilibrate.  The real part gives information about the energy of the mode, i.e. the frequency of the fluctuation.
\\ \indent Motivated by the work of \cite{K}, we are going to consider higher derivative corrections to the magnetic black brane  metric and to tensor QNMs in a coupling corrected magnetic black brane background. The propagator of these perturbations $h_{xy}$ is dual to the two-point function of the $xy$ component of the boundary stress energy tensor. Our numerical analysis gave a well converging result for the lowest $\alpha'$-corrected QNM frequeny, which is the most interesting regarding the equilibration of a QGP in a strong background field. The numerical errors of the $\alpha'$-corrections to the following QNMs were too large to give  results, whose precision exceeds their rough size.\\ \indent  
On the one hand we want to  study how the late time behaviour of the QGP changes, if we consider it to be in a strong magnetic  field, as produced for a very a short time during actual heavy ion collisions, and include higher $\alpha'$ corrections, to leave the $\lambda \to \infty$ limit. On the other hand this analysis also has a more abstract application: So far, we don't have a satisfying dual  theory, that describes QCD.  The most prominent AdS/CFT duality allows us to non-perturbatively compute quantities in a conformal field theory, with $N \to \infty$ and $\lambda=g_{\text{YM}}^2N \to \infty$. Whereas QCD has a finite coupling, a finite  $N=3$ and is not conformally invariant. Apart from (bottom-up-)  modeling, one should try everything that is feasible on the gravity side, to bring the dual field theory closer to QCD in a top-down fashion. This includes the computation of finite coupling corrections, $1/N$ corrections, breaking the scale invariance e.g. by introducing a magnetic background field, where the metric ansatz describing this setting can be deduced from a solution to $10$D SUGRA, or several of the above simultaneously.
\\\indent 
In the limit $\lambda \to \infty$ the holographic description of a QGP in a magnetic background field  was  realized in \cite{K} by considering a Einstein-Maxwell-Chern-Simons theory. That this setting describes the physical properties of the real $SU(3)$ QGP at least qualitatively was shown in \cite{En}. In this work we will show and also need that the ansatz chosen in 
\cite{K} can be derived  from a specific solution to SUGRA living in $10$ dimensions 
(see \cite{emp}). This allows us to determine $\alpha'^3$-corrections first to the metric of a magnetic black brane, where the magnetic background field back-reacts to the metric, and afterwards to QNM fluctuations around this specific solution. We are going to give a mathematical proof of a  prescription, which was found in  \cite{p1}, to handle higher derivative correction to the five form $F_5$ in the presence of gauge fields for the specific case of a constant background field.  The higher derivative corrections to the QNM frequencies and the metric will be computed numerically using pseudo-spectral methods.
\section{Reviewing magnetic black branes in the $\lambda \to \infty$ limit}
\label{section1}
In this chapter we give a review of calculations and results of \cite{K} and present the computations in a way, that makes it more intuitive to extend them to the finite $\lambda$ case.
\\ \indent
 The  action in five dimensions, which is the starting point of the $\lambda=\infty$ calculations of \cite{K} reads
\begin{equation}
S=      \frac{1}{2 \kappa} \int d^5x \sqrt{-g_5} \big[(R_5-2\Lambda)- F_{\mu \nu} F^{\mu \nu} \big],
\label{action1}
\end{equation}
where $\kappa=\frac{1}{8 \pi G_{N}}$ with Newton constant $G_N $, $\Lambda = -6$, $g_5$ is the determinant of the $5$-dimensional metric and $F_{\mu \nu}= \partial_\mu A_\nu-\partial_\nu A_\mu$ for a gauge field  $A_\mu$. As shown by the authors of \cite{T} the five sphere metric components depend on the radial coordinate of the AdS-space, if we consider $\alpha'$ corrections. This will, of course, stay true, when we include a strong magnetic background field with back-reaction on the geometry. Therefore it is advisable to return to the $10$-dimensional type IIB SUGRA action, from which (\ref{action1}) can be derived by integrating out the five sphere coordinates.
\begin{equation}
S_{10} = \frac{1}{2 \kappa} \int d^{10}x \sqrt{-g_{10}} \bigg[R_{10}-\frac{1}{4 \times 5!} F_5^2 \bigg],
\label{action2}
\end{equation} The metric ansatz for a constant magnetic background field with field strength tensor $F_{xy}=b r_h^2=-F_{yx}=\text{const.}$ is given by
  \begin{equation}
  ds_{10}^2= ds_{\text{AdS}}^2+L(u)^2\sum_{i=1}^3 \big(d\mu_i^2+\mu_i^2 (d\phi_i+\frac{2}{\sqrt{3}} A_\mu dx^\mu)^2 \big),
  \label{metricc}
  \end{equation}
  where
  \begin{align}
ds_{AdS}^2= &-r_h^2 U(u)dt^2+ \tilde U(u) du^2+ r_h^2 e^{2 V(u)}(dx^2+dy^2)+r_h^2 e^{2 W(u)} dz^2,
\label{metric}
\end{align}
with $A_y=r_h^2 x b$, $A_\mu=0$ for other directions and $u =\frac{r_h^2}{r^2}$. The five-sphere $S_5$ is described by the coordinates $y_1,\dots,y_5$ with
\begin{align}
\mu_1=\sin(y_1),\text{ }\mu_2=\cos(y_1)\sin(y_2),\text{ }\mu_2=\cos(y_1)\cos(y_2),\text{ }\phi_1=y_3,\text{ }\phi_2=y_4,\text{ }\phi_3=y_5.
\end{align} We have chosen the $xy$-direction of the field strength tensor to be $r_h^2 b$, such that $b$ coincides with the corresponding magnetic field strength parameter  chosen in 
\cite{K}. In the following we are going to set $r_h=1$, which corresponds to a rescaling of the coordinates. Reintroducing $r_h$ in the final differential equations for e.g. tensor fluctuations by 
$\frac{\omega}{2} \to \hat{\omega}=\frac{\omega}{2r_h}$ and $\frac{q}{2} \to \hat q=\frac{q}{2r_h}$, where $\omega$ and $q$ are the frequency and the momentum of the mode  corresponds to a rescaling  to get the original form of the metric (\ref{metric}), (\ref{metricc}). The relation between $b$ and the physical magnetic field is given by \cite{K}
\begin{equation}
B =\frac{b}{v},
\end{equation}
where the constant $v$ can be computed from the near boundary metric. \\ \indent
The self dual solution to the EoMs for the five form components
\begin{equation}
d * F_5=0
\end{equation}
is
\begin{equation}
(F_5^0)^{el}=-\frac{4}{L(u)^5} \epsilon_{AdS}, \hspace{0.5 cm} (F_5^1)^{el}=\frac{1}{\sqrt{3}L(u)}\sum_{i=1}^3 d(\mu_i^2) \wedge d\phi_i \wedge \bar{*}F_2,
\label{F5}
\end{equation}
and 
\begin{equation}
F_5=(1+*)((F_5^0)^{el}+(F_5^1)^{el}),
\label{F50}
\end{equation} 
with $F_2=d A$. Here and henceforth we call $F_5^{el}$ the electric part of the five form and its Hodge dual $F_5^{mag}=*F_5^{el}$ the magnetic part.\footnote{Admittedly this is a misleading notation, since both the electric part and the magnetic part of the five form depend on the magnetic background field. We use this nomenclature to be consistent with the literature.} In the $\lambda=\infty$ case the action (\ref{action1}) is the result of this setup in $10$ dimensions with $L(u)=1$. The factor $\frac{1}{L(u)}$ in front of the second term in (\ref{F5}) was not omitted, although $L(u)=1$ in this order in $\alpha'$, since later on we will need an expression for $F_5$  for which 
\begin{equation}
dF^{mag}=d*F^{el}=0
\end{equation}
for arbitrary $L(u)$ and (\ref{F5}) does the job.  The Einstein equations, or equivalently the differential equations obtained by varying action (\ref{action1}) with respect to $U$, $\tilde U$, $W$, $L$ and $V$ are given by
\begin{align}
0=&b^2 L(u)^{12}+2 b^2 L(u)^4+30 L(u)^8 e^{4 V(u)}(4 u^3 U(u) L'(u)^2-1)+30 u^2 L(u)^9 e^{4 V(u)}\nonumber\\ &(u L'(u) U'(u)+U(u) (2 u L''(u)+L'(u) (4 u V'(u)+2
   u W'(u)+3)))+6 u^2 L(u)^{10}\nonumber\\ & e^{4 V(u)} (u U'(u) (2 V'(u)+W'(u))+U(u) (4 u V''(u)+V'(u) (4 u W'(u)+6)\nonumber\\ &+6 u V'(u)^2+2 u
   W''(u)+2 u W'(u)^2+3 W'(u)))+12 e^{4 V(u)}
\end{align}
\begin{align}
0=&b^2 L(u)^{12}+2 b^2 L(u)^4+30 u^3 L(u)^9 e^{4 V(u)} L'(u) (U'(u)+2 U(u) (2 V'(u)+W'(u)))\nonumber\\ & +30 L(u)^8 e^{4 V(u)} (4 u^3 U(u) L'(u)^2-1)+6 u^3
   L(u)^{10} e^{4 V(u)} (U'(u) (2 V'(u)+ \nonumber\\ &W'(u)) +2 U(u) V'(u) (V'(u)+2 W'(u)))+12 e^{4 V(u)}
\end{align}
\begin{align}
0=&7 b^2 L(u)^{12}-2 b^2 L(u)^4+90 L(u)^8 e^{4 V(u)} (4 u^3 U(u) L'(u)^2-1)+120 u^2 L(u)^9 e^{4 V(u)}\nonumber\\ &  (2 u L'(u) U'(u)+U(u) (2 u L''(u)+L'(u) (4 u
   V'(u)+2 u W'(u)+3)))+15 u^2 \nonumber\\ & L(u)^{10} e^{4 V(u)} (2 (u U''(u)+U(u) (4 u V''(u)+V'(u) (4 u W'(u)+6)+6 u V'(u)^2\nonumber\\ & +2 u W''(u)+2 u
   W'(u)^2+3 W'(u)))+U'(u) (8 u V'(u)+4 u W'(u)+3))\nonumber\\ & -60 e^{4 V(u)}
\end{align}
\begin{align}
0=&b^2 L(u)^{12}+2 b^2 L(u)^4-30 L(u)^8 e^{4 V(u)} (4 u^3 U(u) L'(u)^2-1)-30 u^2 L(u)^9 e^{4 V(u)} \nonumber\\ & (2 u L'(u) U'(u)+U(u) (2 u L''(u)+L'(u) (2 u
   V'(u)+2 u W'(u)+3)))-3 u^2\nonumber\\ &  L(u)^{10} e^{4 V(u)} (2 u U''(u)+U'(u) (4 u V'(u)+4 u W'(u)+3)+U(u) (4 u
   (V''(u)\nonumber\\ & +W''(u))+V'(u) (4 u W'(u)+6)+4 u V'(u)^2+4 u W'(u)^2+6 W'(u)))\nonumber\\ & -12 e^{4 V(u)}
\end{align}
\begin{align}
0=&b^2 L(u)^{12}+2 b^2 L(u)^4+30 L(u)^8 e^{4 V(u)} (4 u^3 U(u) L'(u)^2-1)+30 u^2 L(u)^9 e^{4 V(u)} \nonumber\\ & (2 u L'(u) U'(u)+U(u) (2 u L''(u)+L'(u) (4 u
   V'(u)+3)))+3 u^2 L(u)^{10} e^{4 V(u)}\nonumber\\ &  (2 (u U''(u)+U(u) (4 u V''(u)+6 u V'(u)^2+6 V'(u)))+U'(u)(8 u
   V'(u)+3))\nonumber\\ & +12 e^{4 V(u)},
\end{align}
where we already inserted (\ref{relU}). The ansatz to solve these can be written as
\begin{align}
U(u)&=u_0+u_1(1-u)+u_2 (1-u)^2+\dots\\ 
\tilde{U}(u)&=\frac{1}{4 u^3 U(u)}\label{relU}\\
V(u)&=v_0+v_1(1-u)+v_2(1-u)^2+\dots\\
W(u)&=w_0+w_1(1-u)+w_2(1-u)^2+\dots\\
L(u)&=l_0+l_1(1-u)+l_2(1-u)^2+\dots.
\end{align}
As said above we have for $\lambda=\infty$ that $L(u)=1$, which can be seen from the form of the solution below.
Furthermore we use the freedom to set $u_0=0$, in order to obtain a blackening factor and set $v_0 = w_0 =0$, which can be achieved by rescaling.  As pointed out by \cite{K} $u_1$ is linked to the temperature of the system. In practical calculations we can set $u_1=2$ to give a Schwartzschild black hole for $b \to 0$, which together with our metric ansatz (\ref{metric}) links the temperature to the horizon radius $r_h$. Solving this system of differential equations near the horizon gives
\begin{align}
u_2=&-\frac{-b^2 l_0^{12}-4 b^2  l_0^4-9  l_0^{10} u_1+30  l_0^8-24}{12  l_0^{10}}\\
v_1=&-\frac{b^2 l_0^{12}+b^2 l_0^4-6}{6 l_0^{10} u_1}\\
w_1=&-\frac{-b^2 l_0^4-6}{6 l_0^{10} u_1}\\
l_1=&-\frac{-b^2 l_0^{12}+b^2 l_0^4-30 l_0^8+30}{30 l_0^9 u_1}
\end{align}
The next order term in this expansion is given in the Appendix \ref{expansion}.

Setting $l_0=1$ gives the same expansion as in \cite{K}, with $l_i=0$ for all $i >0$. What we are after is a solution in order $\mathcal{O}(\gamma^0)$ with minimal error on a sufficiently large $u$-interval
 $[l,k] \subset [0,1]$. The solution for the geometry in  order $\mathcal{O}(\gamma^0)$ is obtained by an expansion around the horizon to high order after a near-boundary expansion to low order. 
 After setting $b=\frac{5}{4}$, which corresponds to a physical strong background field of $\mathcal{B}=34.4555 T^2$ 
\cite{K} in the limit $\lambda \to \infty$ 
\footnote{The relation between $b$ and $\mathcal{B}$ deduced from the trace anomaly of the stress energy tensor might get finite coupling corrections, too. Since our focus is on how QNMs behave for large magnetic background fields, without the need to prioritize a precise value for $\mathcal{B}$, we will carry out the calculation including coupling corrections also  with the choice  $b=\frac{5}{4}$, while  stressing that this only approximately  corresponds to the $\lambda \to \infty$ result $\mathcal{B}\approx 34.5 T^2$.}, and introducing the new functions
\begin{align}
U(u)&=\frac{1+u^2 U^{0}(u)}{u}\\
V(u)&=\frac{1}{2}\log(\frac{V^{0}(u)}{u})\\
W(u)&=\frac{1}{2}\log(\frac{W^{0}(u)}{u})\\
L(u)&=1\\
\tilde{U}(u)&=\frac{1}{4 u^3 U^0(u)}
\end{align}
we expand $U^{0}(u)$, $V^{0}(u)$ and $W^{0}(u)$ in $1-u$ and solve the resulting equations order by order up to order $260$ in $(1-u)$.
\section{Higher derivative corrections} 
In the following we will include 't Hooft coupling corrections in our calculations. We start again from the  action in $10$ dimensions, but now with  $\alpha'^3$-correction terms determined in  \cite{Paulos:2008tn}. These terms can be schematically written as 
\begin{equation}
S^\gamma_{10}= \frac{1}{2\kappa} \int d^{10}x \sqrt{\vert g_{10}\vert} \bigg[C^4+C^3 \mathcal{T}+C^2  \mathcal{T}^2 +C \mathcal{T}^3+\mathcal{T}^4\bigg],
\label{gammaaction}
\end{equation}
where we have ignored a factor containing the exponential of the dilaton field, which is $0$  for $\lambda \to \infty$, and written the contractions between the tensors $C$ and $\mathcal{T}$, which will be defined below, as products. The quantity $\gamma$ is defined as $\gamma = \frac{\zeta(3)\lambda^{-\frac{3}{2}}}{8}$ and is thus proportional to $\alpha'^3$. Correction terms to the type IIb SUGRA action of order $\alpha'$ and $\alpha'^2$ vanish.
The action we work with in the following can be written as
\begin{equation}
S=S_{10}+\gamma S^{\gamma}_{10}+\mathcal{O}(\gamma^{\frac{4}{3}}).
\label{actiongamma}
\end{equation}
 $C_{abcd}$ is the Weyl tensor  of the ten dimensional manifold 
 and $\mathcal{T}$ is given by
\begin{equation}
\mathcal{T}_{abcdef}= i \nabla_a F_{bcdef}^++\frac{1}{16} \big(F_{abcmn}^+ F_{def}^{+ mn}-3 F_{abfmn}^+ F_{dec}^{+ mn}\big),
\label{Ttensor}
\end{equation}
with antisymmetrized indices $a,b,c$ and $d,e,f$ and symmetrized with respect to the interchange of $(a,b,c) \leftrightarrow (d,e,f)$ \cite{Paulos:2008tn}. Here $F^{+}$ is the self dual part of the $F_5$ ansatz or $F^+=\frac{1}{2}(1+*)F_5$ (working with Lorentzian signature ensures that this part of $F_5$ exists).  Using the notation in \cite{Paulos:2008tn} we write
\begin{equation}
\gamma W = \gamma\bigg[C^4+C^3 \mathcal{T}+C^2  \mathcal{T}^2 +C \mathcal{T}^3+\mathcal{T}^4\bigg]
\end{equation} with 
\begin{equation}
\gamma W = \frac{\gamma}{86016} \sum_{i=1}^{20} n_i M_i
\end{equation}
and
\begin{align}
\left(n_i\right)_{i=1,\dots, 20}=&(-43008,86016,129024,30240,7392,-4032,-4032,
-118272,\nonumber \\ &-26880,112896,-96768,1344,-12096,-48384,24192,2386,\nonumber \\ &-3669,
-1296,10368,2688 )
\end{align}
as well as
\begin{align}
(M_i)_{i=1,\dots,20}= &(C_{abcd} C^{abef}C^{c} \hspace{0.01cm}_{egh}C^{dg}\hspace{0.01cm}_f\hspace{0.01cm}^h, C_{abcd}C^{aecf} C^{bg}\hspace{0.01cm}_{eh}C^d\hspace{0.01cm}_{gf}
\hspace{0.01cm}^h, \\ \nonumber & C_{abcd}C^a\hspace{0.01cm}_e\hspace{0.01cm}^f
\hspace{0.01cm}_g C^b\hspace{0.01cm}_{fhi}\mathcal{T}^{cdeghi},C_{abc}
\hspace{0.01cm}^d C^{abc}\hspace{0.01cm}_e\mathcal{T}_{dfghij}
\mathcal{T}^{efhgij},\nonumber \\ &
C_a\hspace{0.01cm}^{bcd}C^a\hspace{0.01cm}_{bef}
\mathcal{T}_{cdghij}\mathcal{T}^{efghij},
C_a\hspace{0.01cm}^{bc}\hspace{0.01cm}_dC^{ae}\hspace{0.01cm}_{cf}
\mathcal{T}_{beghij}\mathcal{T}^{dfghij}\nonumber  \\ &
C_a\hspace{0.01cm}^{bcd} C^a\hspace{0.01cm}_{ecf}\mathcal{T}_{bghdij}
\mathcal{T}^{eghfij},
C_a\hspace{0.01cm}^{bc}\hspace{0.01cm}_{d}
C^{ae}\hspace{0.01cm}_{fg}\mathcal{T}_{bcehij}
\mathcal{T}^{dfhgij}, \nonumber \\ &C_a\hspace{0.01cm}^{bc}\hspace{0.01cm}_{d}
C^{ae}\hspace{0.01cm}_{fg}\mathcal{T}_{bcehij}
\mathcal{T}^{dhifgj},C_a\hspace{0.01cm}^{bc}\hspace{0.01cm}_{d}
C^{a}\hspace{0.01cm}_e\hspace{0.01cm}^{f}\hspace{0.01cm}_g
\mathcal{T}_{bcfhij}
\mathcal{T}^{dehgij}, \nonumber \\ &C_a\hspace{0.01cm}^{bc}\hspace{0.01cm}_{d}
C^{ae}\hspace{0.01cm}_{fg}\mathcal{T}_{bcheij}
\mathcal{T}^{dfhgij}, C^{a b c d} \mathcal{T}_{abefgh}\mathcal{T}_{cd}\hspace{0.01cm}^{eijk}
\mathcal{T}^{fgh}\hspace{0.01cm}_{ijk},\nonumber \\ & C^{a b c d} \mathcal{T}_{abefgh}\mathcal{T}_{cd}\hspace{0.01cm}^{fijk}
\mathcal{T}^{egh}\hspace{0.01cm}_{ijk}, C^{a b c d} \mathcal{T}_{abefgh}\mathcal{T}_{cd}\hspace{0.01cm}^{fijk}
\mathcal{T}^{eg}\hspace{0.01cm}_i\hspace{0.01cm}^h
\hspace{0.01cm}_{jk}, \nonumber  \\ & C^{a b c d} \mathcal{T}_{abefgh}\mathcal{T}_{c}\hspace{0.01cm}^{efijk}
\mathcal{T}_{d}\hspace{0.01cm}^{gh}
\hspace{0.01cm}_{ijk},\mathcal{T}_{abcdef}\mathcal{T}^{abcdgh}\mathcal{T}^e
\hspace{0.01cm}_{gijkl}\mathcal{T}^{fij}\hspace{0.01cm}_h
\hspace{0.01cm}^{kl}, \nonumber  \\ &\mathcal{T}_{abcdef}\mathcal{T}^{abcghi}\mathcal{T}^{de}
\hspace{0.01cm}_{jg}\hspace{0.01cm}^{kl}\mathcal{T}^{f}
\hspace{0.01cm}_{hki}
\hspace{0.01cm}^{j}\hspace{0.01cm}_l,\mathcal{T}_{abcdef}\mathcal{T}^{abcghi}\mathcal{T}^{d}
\hspace{0.01cm}_{gj}\hspace{0.01cm}^{ekl}\mathcal{T}^{f}
\hspace{0.01cm}_{h}
\hspace{0.01cm}^{j}\hspace{0.01cm}_{ikl}\nonumber \\ &
\mathcal{T}_{abcdef}\mathcal{T}^{abcghi}\mathcal{T}^{d}
\hspace{0.01cm}_{gj}\hspace{0.01cm}^{ekl}\mathcal{T}^{f}
\hspace{0.01cm}_{hki}
\hspace{0.01cm}^{j}\hspace{0.01cm}_l, \mathcal{T}_{abcdef}\mathcal{T}^{aghdij}\mathcal{T}^{b}
\hspace{0.01cm}_{gk}\hspace{0.01cm}^e\hspace{0.01cm}_{il}\mathcal{T}^{c}\hspace{0.01cm}_{h}
\hspace{0.01cm}^{kf}\hspace{0.01cm}_{j}\hspace{0.01cm}^l
).
\end{align}
The higher derivative corrected EoM for the five form is given by 
\begin{equation}
d \bigg(*F_5-*\frac{2\gamma}{\sqrt{-g}}\frac{\delta \mathcal{W}}{\delta F_5} \bigg)=0,
\label{EoMF5}
\end{equation}
which yields 
\begin{equation}
F_5=*F_5-*\frac{2\gamma}{\sqrt{-g}}\frac{\delta \mathcal{W}}{\delta F_5},
\label{EqF5}
\end{equation}
where we set
\begin{equation}
\frac{\delta \mathcal{W}}{\delta F_5}:=2 \kappa\frac{\delta S_{10}^\gamma}{\delta F_5}.
\end{equation}
\subsection{A helpful prescription and its mathematical proof}
\label{theorem}
In this section we claim and proof the validity of the following prescription, which will facilitate our calculation noticeably. It is equivalent to strictly applying the variational principle, treating both the four form components and the metric as independent fields and solve the resulting system of highly coupled, finite coupling corrected  differential equations simultaneously including the back-reaction of a strong background field:\\ \indent Solve the equation of motion for $F_5$ in the lowest order in $\alpha'$ for a strong background field, such that it depends on  the metric components of the ansatz made in (\ref{metricc}, \ref{metric}) (which we allow to be of order $\mathcal{O}(\gamma)$)
and choose the $L(u)$-factor  of the components of the electric part of the five form in such a way that 
\begin{equation}
d F^{mag}=d * F^{el}=\mathcal{O}(\gamma^2)=d  F^{el}.
\label{condition1}
\end{equation}
Now replace the $F_5^2$ term in the action with 2 times $(F^{mag})^2$ and insert  $F_5$ as given in (\ref{F5}, \ref{F50}), which depends on metric components, that still have to be determined, into the higher derivative part of the action. The resulting action only depends on the absolute value of the $z$-component of the magnetic background field $b$ and the metric, whose solution in order $\mathcal{O}(\gamma)$ will be determined by solving the system of differential equations obtained by varying this effective action with respect to $g^{\mu \nu}$.\footnote{Observing that $F_5 \wedge * F_5 =F_5\wedge \Big(*\frac{2 \gamma}{\sqrt{-g}}\big(\frac{\delta \mathcal{W}}{\delta F_5}\big)\Big)$ is the starting point of generalizing the following proof to arbitrary gauge fields.}\\ \indent We justify this claim with the following proof, where we work with the metric ansatz given in (\ref{metricc}), (\ref{metric}).
\begin{lemma}In order $\mathcal{O}(x^0)$ the magnetic parts of the five form don't get any $\gamma$-corrections, except for those coming from the finite $\lambda$ correction to the metric. The non-trivial higher derivative corrections to the electric parts of $F_5$ (i.e. the finite $\lambda$ terms, which are not caused by corrections to the metric, the $\mathcal{O}(\gamma^0)$ solution of $F_5$ depends on) are given by the respective directions of \begin{equation}
\frac{2 \gamma}{\sqrt{-g}}\bigg(\frac{\delta \mathcal{W}}{\delta F_5}\bigg).
\end{equation}
\label{lemma1}
\end{lemma}
\begin{beweis}
Let us first focus on the $tuzy_3$-component of $C_4$. In order $\mathcal{O}(x^0)$ the diagram describing the system of differential equations it appears in, derived from (\ref{EoMF5}), is given by
\begin{equation} 
\begin{xy}
  \xymatrix{
 (C_4)_{tuzy_3} \ar[rd]^d\ar[rdd]^d&   &   & (d*F_5)_{uxyy_2y_4y_5} \\
         & (F_5)_{ tuzy_1y_3}\ar[r]^{*} &   (*F_5)_{x y y_2 y_4 y_5} \ar[r]^d \ar[ru]^d  & (d*F_5)_{x y y_1 y_2 y_4 y_5}\\
         &   (F_5)_{tuxzy_3}\ar[r]^{*} &   (*F_5)_{y y_1 y_2 y_4 y_5}\ar[ru]^d \ar[r]^d &  (d*F_5)_{uy y_1 y_2 y_4 y_5}  \\ 
  }
  \label{diag1}
\end{xy}
\end{equation}
where the right hand side has to be equal to the corresponding directions of \begin{equation}
d*\bigg(\frac{2 \gamma}{\sqrt{-g}}\frac{\delta \mathcal{W}}{\delta F_5}\bigg).
\end{equation}
In order $\mathcal{O}(x^0)$ there are no other contributions from $C_4$ to the right hand side of the diagram.  From diagram (\ref{diag1}) we can derive that modulo  terms, which are independent of $u$, the following equations hold
\begin{equation}
(F_5)_{ tuzy_1y_3}=\frac{2 \gamma}{\sqrt{-g}}\bigg(\frac{\delta \mathcal{W}}{\delta F_5}\bigg)_{tuzy_1y_3}+(\tilde F_5)_{ tuzy_1y_3}+\mathcal{O}(x^1),
\label{F51}
\end{equation}
\begin{equation}
(F_5)_{tuxzy_3}= \frac{2 \gamma}{\sqrt{-g}}\bigg(\frac{\delta \mathcal{W}}{\delta F_5}\bigg)_{tuxzy_3}+(\tilde F_5)_{ tuxzy_3}+\mathcal{O}(x^1),
\label{F52}
\end{equation}
where $\tilde F_5$ describes the five form solution, depending on arbitrary metric components,
shown in  (\ref{F5}) and (\ref{F50}). Notice that we already used relation (\ref{condition1}) (where $\tilde F_5$ corresponds to $ F_5$ there) when deducing the solutions  (\ref{F5}) and (\ref{F50}).
The $u$-independent terms, which in theory could be added to equation (\ref{F51}) and (\ref{F52}), if they don't corrupt the diagram dual to (\ref{diag1}), can be gauged away, since they correspond to terms in $C_4$, which only give contributions to $(d*F_5)_{x y y_1 y_2 y_4 y_5}$. 
Very similar calculations provide analogous relations for the 
\begin{equation}
tuzy_1y_4,\hspace{0.1cm}tuzy_1y_5,\hspace{0.1cm}tuzy_2y_4,\hspace{0.1cm}tuzy_2y_5,\hspace{0.1cm}tuxzy_3,\hspace{0.1cm}tuxzy_4,\hspace{0.1cm}tuxzy_5\hspace{0.1cm}-
\end{equation}
directions  of the five form. Considering now equation (\ref{EqF5}) proves this lemma for those directions of the five form, which in the $\lambda=\infty$ limit are of order $\mathcal{O}(b^1)$ or higher. The analogous diagram for the $txyz$  direction of the four form $C_4$ is even easier and gives results analogous to (\ref{F51}), such that Lemma \ref{lemma1} follows  by again applying relation (\ref{EqF5}).
\end{beweis}
\begin{lemma}
\label{lemma2}
The magnetic parts of the five form components in (\ref{F5}) with arbitrary $L(u)$, with lower indices and the electric parts of the five form components in (\ref{F5}) with arbitrary $L(u)$, with upper indices times $\sqrt{-g}$ are independent of $u$. 
\end{lemma}
\begin{beweis}
This claim follows by carefully inspecting the magnetic part $F_5^{mag}=*F^{el}_5$ of  $F_5$ given in (\ref{F5}) and (\ref{F50})  and by using the self duality of this five form.
\end{beweis}
\begin{bemerkung}
\label{lemma3}
The magnetic parts of the five form components in (\ref{F5}) with arbitrary $L(u)$, with lower indices and the electric parts of the five form components in (\ref{F5}) with arbitrary $L(u)$, with upper indices times $\sqrt{-g}$ are actually independent of the AdS-part of the metric and independent of $L(u)$ if we choose the $L(u)$ factor of the magnetic part of the five form so that (\ref{condition1}) holds. 
\end{bemerkung}
\begin{lemma}
For any five form, which doesn't depend on derivatives of a metric component $X \in \{g_{ \mu \nu}\}_{\mu \nu \in \{1,\dots,10\}}$, we have 
\begin{equation}
\frac{\partial \partial_u (F_5)_{abcde}}{\partial\partial_u X}=\frac{\partial (F_5)_{abcde}}{\partial X}
\end{equation}
for all directions $abcde$.
\label{lemma4}
\end{lemma}
\begin{beweis}
Let $\{X_i\}_{i \in I}$ be equal to the set $\{g_{ \mu \nu}\}_{\mu \nu \in \{1,\dots,10\}}$ and let $X_0=X$. Then we have 
\begin{equation}
\frac{\partial \partial_u (F_5)_{abcde}}{\partial\partial_u X}=\frac{\partial}{\partial \partial_u X}\frac{\partial (F_5)_{abcde}}{\partial X_i}\partial_u X_i=\frac{\partial}{\partial \partial_u X}\frac{\partial (F_5)_{abcde}}{\partial X_0}\partial_u X_0=\frac{\partial (F_5)_{abcde}}{\partial X},
\end{equation}
where we made use of the sum convention.
\end{beweis}
\begin{lemma}
\label{lemma5}
For any direction $abcde$ of $F_5$ and any metric component $X$ corresponding to the internal AdS$_5$-space or $L(u)$ we have that
\begin{equation}
\frac{\partial \mathcal{W}}{\partial (F_5)_{abcde}}\frac{\partial (F_5)_{abcde}}{\partial X}+\frac{\partial \mathcal{W}}{\partial \partial_u (F_5)_{abcde}}\frac{\partial \partial_u (F_5)_{abcde}}{\partial X}-\frac{d}{du}\bigg(\frac{\partial \mathcal{W}}{\partial \partial_u (F_5)_{abcde}}\frac{\partial \partial_u (F_5)_{abcde}}{\partial \partial_u X}\bigg)
\label{eq10}
\end{equation} is equal to 
\begin{equation}
\bigg(\frac{\partial \mathcal{W}}{\partial (F_5)_{abcde}}-   \frac{d}{du}\frac{\partial \mathcal{W}}{\partial \partial_u(F_5)_{abcde}}\bigg) \frac{\partial (F_5)_{abcde}}{\partial X}.
\end{equation}
\end{lemma}
\begin{beweis}
The claim follows immediately with Lemma  \ref{lemma4}.
\end{beweis}
\begin{satz}
The prescription given in the introduction of this section is valid.
\label{satz1}
\end{satz}
\begin{beweis} Due to Lemma \ref{lemma1} and due to the fact that the effective action for the metric is not allowed to depend on $x$, because of gauge invariance, the theorem \ref{satz1} holds, if we can show that for any given direction $abcde$, for which the electric part of the five form $F_5$is non-zero, the expression given by $-\gamma$(\ref{eq10})$|_{g \to \mathfrak{g}}$ is the same as
\begin{align}
&\bigg(\frac{\partial}{\partial X}\gamma \frac{\sqrt{-g}}{\sqrt{-\mathfrak{g}}}\bigg(\mathfrak{g}_{a a} \mathfrak{g}_{b b}\mathfrak{g}_{c c} \mathfrak{g}_{d d}\mathfrak{g}_{e e}   \bigg(\frac{\partial \mathcal{W}}{\partial(F_5)_{abcde}}-\frac{d}{du}\frac{\partial \mathcal{W}}{\partial_u\partial(F_5)_{abcde}} \bigg)\bigg|_{g \to \mathfrak{g}} \bigg)g^{aa}g^{bb}g^{cc}g^{dd}g^ {ee} \nonumber\\ &((F_5)_{abcde}|_{g \to \mathfrak{g}})\bigg)\bigg|_{g \to \mathfrak{g}}+\mathcal{O}(\gamma^2)
\end{align}
for $X \in \{g_{ \mu \nu}\}_{\mu \nu \in \{1,\dots,10\}}$ and $\mathfrak{g}$ being the solution for the metric with back-reaction and without higher derivative corrections. The claim now follows immediately by applying Lemma \ref{lemma5} and Lemma \ref{lemma1}, since  comment \ref{lemma3} implies
\begin{equation}
(\partial_X\sqrt{-g} g^{aa}g^{bb}g^{cc}g^{dd}g^ {ee} ) ((F_5^{el})_{abcde})\Big|_{g \to \mathfrak{g}}=-(\sqrt{-g} g^{aa}g^{bb}g^{cc}g^{dd}g^ {ee} ) ((\partial_X F_5^{el})_{abcde})\Big|_{g \to \mathfrak{g}}.
\end{equation}
\end{beweis}
\indent We also can extend the prescription to include tensor fluctuations.
Similar to the case $b=0$ the tensor fluctuations $h_{xy}$ of the back-reacted and coupling corrected geometry don't change the higher derivative corrected solutions of the five form in a non-trivial way. This means the only way the fluctuations $h_{xy}$ perturb the five form is via the AdS-Hodge-dual $\bar *$ in (\ref{F5}). We now show that the prescription given at the beginning of this section can be extended to also include metric fluctuations 
\begin{equation}
ds_{10}+h_{xy} dx dy
\end{equation}
and their treatment. 
\begin{lemma}
The magnetic part of the $\mathcal{O}(\gamma^0)$ components of the five form with lower indices and the electric part with upper indices times $ \sqrt{-g}$ don't depend on $h_{xy}$.
\label{lemma6}
\end{lemma}
\begin{beweis}
Since
\begin{equation}
\frac{\partial}{\partial h_{xy}}| g| \big(g^{xx}g^{yy}-(g^{xy})^2\big)=0
\end{equation}
the Lemma follows immediately.
\end{beweis} 
The proof of the validity of the extension of the prescription is now entirely analogous to the one presented for
theorem \ref{satz1}.
\subsection{An alternative algorithm to compute higher derivative corrections to the AdS-Schwarzschild black hole solution}
\label{alt}
In this chapter we present a way to compute higher derivative corrections to the  AdS-Schwarzschild black hole solution, so in the case $b=0$, on an interval $u=\frac{r^2}{r_h^2} \in [l,k]\subset [0,1]$.
The interval boundaries  $l$ and $k$ have to be chosen sufficiently close to $0$ and $1$. The following procedure can be generalized to the case of a non-vanishing background field with back-reaction on the geometry. In that case we cannot hope to 
be able to determine the higher derivative corrections to the metric analytically. Even a near boundary
and a near horizon analysis of the higher derivative correction terms to the differential equations of the metric with back-reaction of a strong magnetic background field turns out to be extremely difficult. We motivate the computational strategy we are going to apply to determine these corrections to the metric numerically by performing an analogous calculation in the case $b=0$ and show that it delivers the same results (with very small errors) as the analytical solutions first derived in \cite{T}. \\ \indent

Our metric ansatz is of the form (\ref{metricc}), (\ref{metric}), with $V(u)=W(u)$. The differential equations are obtained by varying the action (\ref{action2}) plus (\ref{gammaaction}) with respect to the functions $L(u)$, $V(u)$, $U(u)$ and $\tilde U(u)$.

\begin{figure}
\includegraphics[scale=0.8]{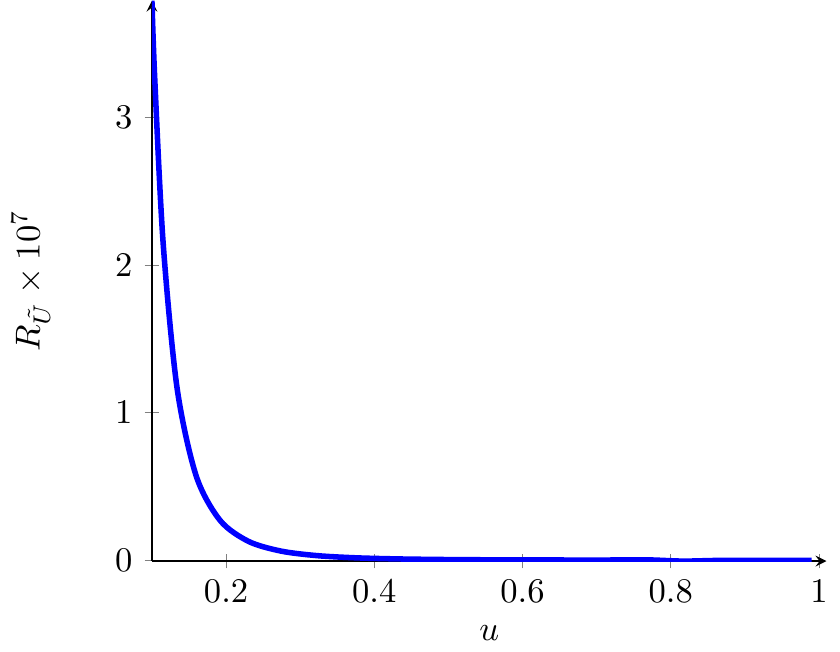}
\includegraphics[scale=0.8]{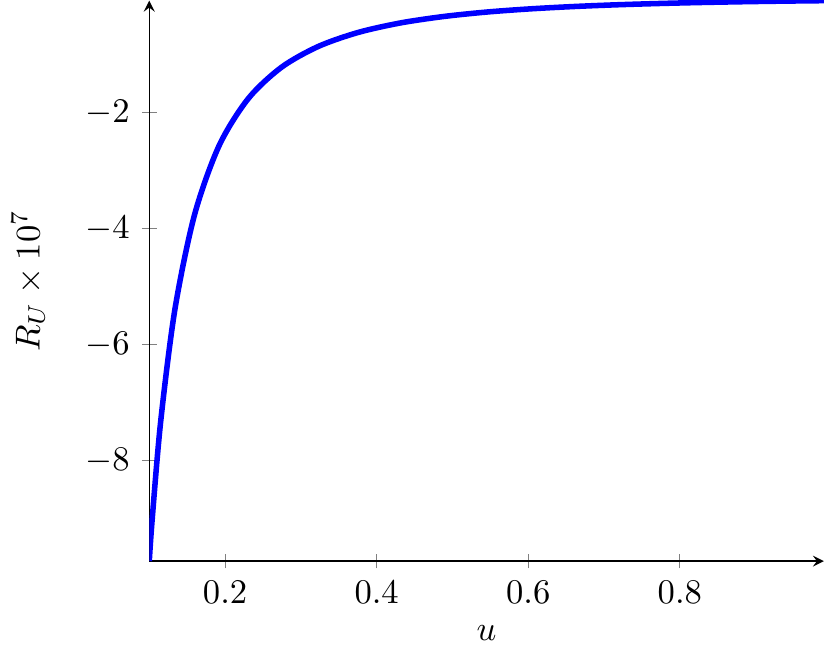}
\caption{Relative error between the analytic solution and the numerical solution $R_{\tilde{U}}$ (left) and $R_U$ (right) as defined in (\ref{reler}), obtained by calculating on a Gauss-Lobatto grid on the interval $[l,k] $, with the choice $d_1=1$, $d_2=0$, $l=0.1$ and $k=0.99$.}
\label{figur1}
\end{figure}

Let now $\mathcal{L}_{10}$ be the action defined in (\ref{action2}) with $F_5^{el}=-\frac{4}{L(u)^5} \epsilon_{\text{AdS}}$. In addition we define
\begin{equation}
\mathcal{L}_{10}^{W}=\sqrt{\vert g_{10}\vert} \bigg[C^4+C^3 \mathcal{T}+C^2  \mathcal{T}^2 +C \mathcal{T}^3+\mathcal{T}^4\bigg],
\end{equation} where the contributions of the $\mathcal{T}$-tensors to the EoM  vanish in the case of absent background  fields $b=0$. We have to solve  the differential equations 
\begin{equation}
\bigg(\frac{\partial}{\partial X(u)}-\frac{d}{du} \frac{\partial}{\partial X'(u)}+\frac{d^2}{du^2}\frac{\partial}{\partial X''(u)}\bigg)\big(\mathcal{L}_{10}+\gamma \mathcal{L}_{10}^W \big)=0,
\label{diff}
\end{equation} with $X(u)\in \{V(u)= W(u),U(u),\tilde{U}(u),L(u)\}$. We choose the ans\"atze
\begin{align}
X(u)= X^0(u)+\gamma X^1(u).
\end{align}
Only the $X^0(u)$ parts are entering the terms
\begin{equation}
\gamma L^{W}_{10}(X)=\bigg(\frac{\partial}{\partial X(u)}-\frac{d}{du} \frac{\partial}{\partial X'(u)}+\frac{d^2}{du^2}\frac{\partial}{\partial X''(u)}\bigg)\gamma \mathcal{L}_{10}^W,
\end{equation} if we want to calculate the coupling corrections up to order $\mathcal{O}(\gamma)$. From the expansion around the horizon and up to order $\mathcal{O}(\gamma)$ of the terms
\begin{equation}
L_{10}(X):=\bigg(\frac{\partial}{\partial X(u)}-\frac{d}{du} \frac{\partial}{\partial X'(u)}+\frac{d^2}{du^2}\frac{\partial}{\partial X''(u)}\bigg) \mathcal{L}_{10}
\end{equation} we can see that $L_{10}^W(X)$ is regular at the horizon for $X(u) \in \{\tilde{U}(u),V(u),L(u)\}$, whereas for $X(u)=U(u)$  it has a pole of at maximum first order at $u=1$.\footnote{Finite coupling corrections don't cause additional poles in the metric.} In the following our aim is to determine the terms $L_{10}^W(X)$. Our strategy will be to calculate the terms 
\begin{equation}
\frac{\partial}{\partial X(u)}\mathcal{L}^W_{10}, \hspace{0.8cm}\frac{\partial}{\partial X'(u)}\mathcal{L}^W_{10}\hspace{0.4 cm} \text{ and } \hspace{0.4 cm} \frac{\partial}{\partial X''(u)}\mathcal{L}^W_{10}
\end{equation} on the rescaled Gauss-Lobatto grid for the $u$-coordinate
\begin{equation}
\frac{l+k}{2}+\frac{l-k}{2}\cos \Big(\frac{\pi n}{M}\Big)_{n \in\{0,\dots, M\}}
\label{grid}
\end{equation}
with $l=0.1$ and $k=0.99$, such that for $u \in [l,k]$ we have
\begin{equation}
x=-\frac{2 u}{l-k}+\frac{l+k}{l-k} \in[-1,1].
\end{equation}
The functions  $U^0(u), \tilde{U}^0(u), V^0(u), W^0(u)$ for a fixed value $b=\frac{5}{4}$ were determined numerically  in section \ref{section1}, in such a way, that the numerical error is negligible on the interval $[l,k]$ on which we have defined our Gauss-Lobatto grid (\ref{grid}). Since we consider the case $b=0$ in this section we perform this calculation with $U^0(u), \tilde{U}^0(u), V^0(u), W^0(u)$ chosen such that (\ref{metric}) is the Schwarzschild black hole metric. The higher derivative corrections will be determined with the help of spectral methods by expanding the ans\"atze  in the following way
\begin{align}
U(u)=& U^0(u)e^{\gamma u^{d_1} \sum_{i=0}^M a^{U,M}_i c^M_i(x u-y)}\label{lasteq1}\\
\tilde{U}(u)=&\tilde{U}^0(u)e^{\gamma u^{d_1} \sum_{i=0}^M a^{\tilde{U},M}_i c^M_i(x u-y)}\\
L(u)=&L^0(u)e^{\gamma u^{d_2} \sum_{i=0}^M a^{L,M}_i c^M_i(x u-y)}\\
V(u)=&W(u)=V^0(u),
\label{lasteq}
\end{align}
 $x=\frac{2}{k-l}$ and $y=\frac{l+k}{k-l}$, $c_i^M$ denotes the $i$-th cardinal function on the grid $\{-\cos(\frac{\pi n}{M})\}_{n \in\{0,\dots, M\}}$ and $a^{\tilde{U},M}_i$, $a^{L,M}_i$, $a^{U,M}_i$ are the respective expansion coefficients. The exact choice of $d_1$ and $d_2$ will be discussed below. The last equation (\ref{lasteq}) follows from the invariance of the metric ansatz under transformations of the form \begin{equation}
u \rightarrow u(\tilde{u})
\end{equation}
to a new radial coordinate $\tilde u$, so that we set $a^{V,M}_i=0$. 
Let $P^\gamma$ be the projection on the first order expansion coefficient  in $\gamma$ of a function $f$, so $P^\gamma f= \frac{\partial}{\partial\gamma}f|_{\gamma \to 0}$, then we have 
\begin{align} 
&P^{\gamma}\big( L_{10}(\tilde{X})+\gamma L_{10}^W(\tilde{X}) \big)\big|_{\{u\to  \frac{y- \cos( \pi n/M)}{x}\}_{n \in \{0,\dots,M\}}}=0
\label{Eing}
\end{align} for each
$
 \tilde{X}\in \{V,U,L,\tilde{U}\} .
$
This can be written as a matrix equation of the form \begin{equation} A \cdot v = \chi,
\end{equation} where for $j \in \{0,1,2\}$, $m \in \{1, \dots, M+1\}$ and $(X_0,X_1,X_2)= (L,U,\tilde{U})$
\begin{align}
&A_{(M+1)j+m,n}= P^{v_n} P^{\gamma}\big( L_{10}(X_j)\big)\big|_{\{u\to  \frac{y- \cos( \pi (m-1)/M)}{x}\}}
 \label{matrix}
\end{align}
is a real $3(M+1) \times 3(M+1)$-matrix. The vector $v$ is given by
\begin{equation}
v_{j(M+1)+m}=a_{m-1}^{X_j,M}
\end{equation}
and finally the $3(M+1)$-vector $\chi$ is
\begin{align}
\chi_{j (M+1)+m}=-\big(L_{10}^W(X_j)|_{\{X(u)\to X^0(u)\}_{X\in \{W,V,L,U,\tilde{U}\}}} \big)\big|_{\{u\to  \frac{y- \cos( \pi (m-1)/M)}{x}\}}.
\end{align}
The resulting system of equations can be solved easily. The equation obtained by inserting $\tilde{X}=V$ in (\ref{Eing}) is and has to be fulfilled by the found solution of (\ref{matrix}).  The near boundary behaviour of the higher derivative corrections to the metric in (\ref{lasteq1})-(\ref{lasteq}) is encoded in the still undetermined exponents $d_1$ and $d_2$. In the original calculation given in \cite{T} the authors choose a specific expansion ansatz to solve the higher derivative corrected EoM for the metric. They showed that the only undetermined expansion coefficient can be swallowed by a rescaling of the time coordinate. 
Simply by rescaling and by the  requirement that the metric on the boundary should be conformally equivalent to the Minkowski metric, one can already reach $0 \leq d_2$ and $1 \leq d_1$. 
The explicit form of  (\ref{lasteq1})-(\ref{lasteq}) with $d_2=4=2d_1 $  follows from a near boundary analysis of the higher derivative corrected Einstein equations. However, we won't make use of this analysis and start the calculation naively with $d_2=0$, $d_1 =1$, since this will also be the strategy in the case $b \neq 0$. Solving the system of equations for the expansion coefficients $\{a_{i}^{X,M}\}_{i\in \{0, \dots, M\}, X\in\{\tilde U,U L\}}$ on the Gauss-Lobatto grid on $[l,k] $ gives results, whose relative errors 
\begin{equation}
R_{X}=\frac{X^{\text{numerical}}_\gamma-X_\gamma^{\text{analytical}}}{X_\gamma^{\text{analytical}}}
\label{reler}
\end{equation}
are displayed in figure (\ref{figur1}) for $M=25$ and for the  first order $\gamma$ corrections to the functions  $U$ and $\tilde{U}$.
\FloatBarrier
The error for $U$ and $\tilde{U}$ are both of order $10^{-7}$,  the relative error for $L$ has a maximal value of $\approx 0.00066$. The solution to the problem of how to improve the numerical precision in a way that can be extended to the $b \neq 0$ case lies in the following observation: \\ \indent
If we choose the interval to be $[l,k]$, with $k=0.99$ as before and $l$ sufficiently large we have to reach a point, where the determinant of the system of equations for the expansion  coefficients of the higher derivative corrections to the metric tends to zero. This is because we thereby admit solutions, which are divergent at the boundary and whose suppression was achieved by choosing $l$ sufficiently small. The same logic applies to the choice of $d_1$ and $d_2$ in (\ref{lasteq1})-(\ref{lasteq}). For a choice of $d_1$ and $d_2$, which is sufficiently far away from the actual near boundary behaviour, the determinant of $A$ in (\ref{matrix}) decreases. We can implicitly determine the near boundary behaviour by minimizing the function
\begin{equation}
\text{min}\Big(\big \{\big((A^{-1})^{numerical}A-1_{3M+3,3M+3}\big)_{a,b}|a,b \in\{1,\dots,3M+3\}\big \}\Big)
\end{equation}
where $(A^{-1})^{numerical}$ is the numerically determined inverse of the matrix in (\ref{matrix}), keeping $M$, $l$, $k$ fixed and only varying $d_1$ and $d_2$. This actually gives $d_1=2=\frac{d_2}{2}$. The  maximal absolute value of the relative error, which again appears for $R_X=R_L$, is now $7.3 \times 10^{-9}$.

\subsection{Calculating higher derivative corrections to the magnetic black brane metric}
\begin{figure}
 \includegraphics[scale=0.8]{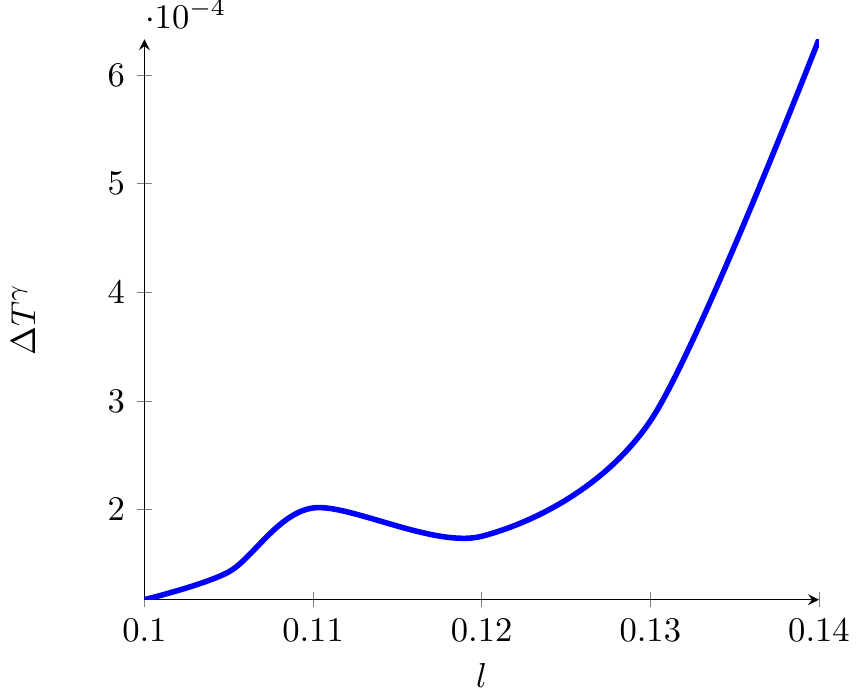}
 \includegraphics[scale=0.8]{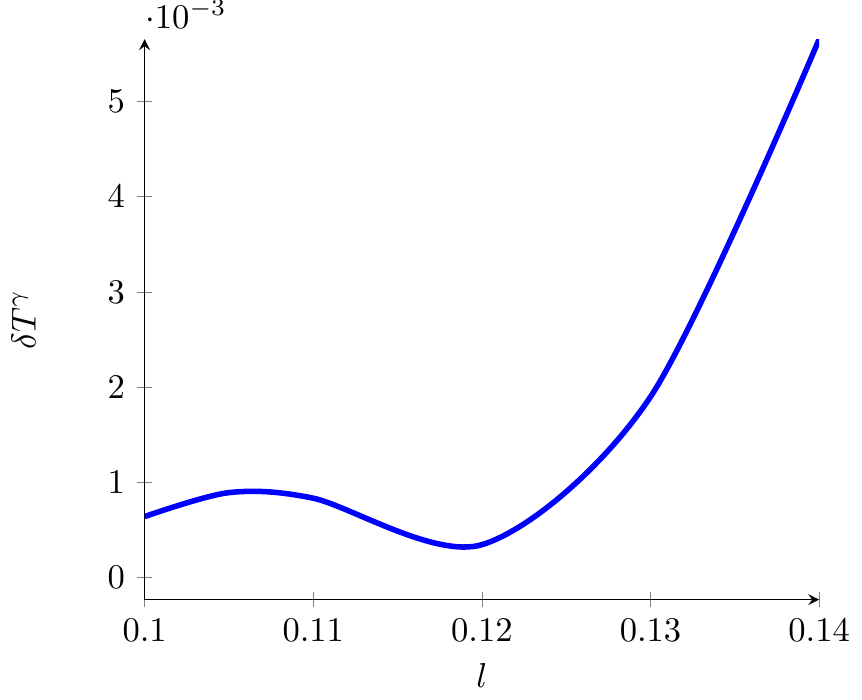}
 \caption{On the left the relative estimated error of the $\gamma$ correction to the temperature averaged over $M$ is plotted for different values of the interval boundary $l$. On the right hand side the function $\delta T ^\gamma(l)$ defined in (\ref{deltaT}) is shown.}
 \label{errorT}
 \end{figure}

In this chapter we are going to generalize techniques derived previously to determine an approximation of  higher derivative corrections to the metric computed in section \ref{section1}. 
First of all we have to use the theorem derived in section \ref{theorem}. We apply the prescription from there to simplify our calculation.  Following this theorem we define the five form $F_5$ in the following way:  
Starting with the fluctuation free electric part and its Hodge dual, we get
\begin{align}
(F_5^{el})^0&= -\frac{4}{L(u)^5}\sqrt{|\det(g_5)|} dt\wedge du \wedge dx\wedge dy \wedge dz\nonumber \\ 
*\Big((F_5^{el})^0\Big) &= 4 \sqrt{\det(g_{S_5})}dy_1\wedge dy_2 \wedge  dy_3 \wedge dy_4 \wedge dy_5+\frac{4}{L(u)^5}\sqrt{|\det(g_{10})|}\sqrt{|\det(g_5)|}\nonumber \\ & \bigg(g_{10}^{t t}g_{10}^{u u}g_{10}^{x x}g_{10}^{y y_3}g_{10}^{z  z} dy_1 \wedge dy_2 \wedge dy \wedge dy_4 \wedge dy_5 +g_{10}^{t t}g_{10}^{u u}g_{10}^{x x}g_{10}^{y y_4}g_{10}^{zz } dy_1 \wedge dy_2   \nonumber \\ &\wedge dy_3\wedge dy \wedge dy_5+g_{10}^{t  t}g_{10}^{u  u}g_{10}^{ x x }g_{10}^{y y_{5}}g_{10}^{ z z} dy_1 \wedge dy_2 \wedge dy_3 \wedge dy_4 \wedge dy\bigg).
\end{align}
The electric components of the five form including the gauge field $A_y = bx$ is explicitly given by
\begin{align}
(F_5^{el})^1&= \frac{2 b}{ \sqrt{3}L(u)}\sqrt{|\det(g_5)|}g_5^{xx} g_5^{yy }\big(\sin(y_1)\cos(y_1)dt \wedge du \wedge dz \wedge dy_1 \wedge dy_3 +\nonumber \\ & \cos(y_1)^2\sin(y_2)\cos(y_2)dt \wedge du \wedge dz \wedge dy_2 \wedge dy_4-\cos(y_1)\sin(y_1)\sin(y_2)^2dt \nonumber \\ &\wedge du \wedge dz \wedge dy_1 \wedge dy_4-\cos(y_1)\sin(y_1)\cos(y_2)^2dt \wedge du \wedge dz \wedge dy_1 \wedge dy_5\nonumber \\ &-\cos(y_2)\sin(y_2)\cos(y_1)^2dt \wedge du \wedge dz \wedge dy_2 \wedge dy_5
 \big),
\end{align}
while its Hodge dual  simplifies to
\begin{align}
*\Big((F_5^{el})^1\Big)&= -\frac{2 b }{ \sqrt{3}}L(u)^4\sqrt{\det(g_{S_5})}\big(\sin(y_1)\cos(y_1)g_{10}^{y_1y_1}(g_{10}^{y_3 y_3} -\sin(y_2)^2g_{10}^{y_4 y_3} - \nonumber\\&\cos(y_1)^2g_{10}^{y_5y_3})dx \wedge dy \wedge dy_2 \wedge dy_5 \wedge dy_4  +\cos(y_1)^2\sin(y_2)\cos(y_2)\times \nonumber \\ & g_{10}^{y_2 y_2}(g_{10}^{y_4 y_4} -g_{10}^{y_5 y_4})dx \wedge dy \wedge dy_1 \wedge dy_5 \wedge dy_3-\sin(y_1)\cos(y_1)g_{10}^{y_1 y_1} \nonumber \\ &(\sin(y_2)^2g_{10}^{y_4 y_4}-g_{10}^{y_3y_4}+g_{10}^{y_5 y_4}\cos(y_2)^2) dx \wedge dy \wedge dy_2 \wedge dy_3 \wedge dy_5- \nonumber \\ &\cos(y_1) \sin(y_1)g_{10}^{y_1 y_1}(\cos(y_2)^2g_{10}^{y_5 y_5} -g_{10}^{y_3 y_5}+g_{10}^{y_4 y_5}\sin(y_2)^2)dx \wedge dy \wedge dy_2  \nonumber \\ &\wedge dy_4 \wedge dy_3-\cos(y_2) \sin(y_2)g_{10}^{y_2 y_2}(g_{10}^{y_5 y_5}\cos(y_1)^2+\sin(y_1)^2g_{10}^{y_4y_5} )dx \wedge dy\nonumber \\ & \wedge dy_1 \wedge dy_3 \wedge dy_4  -\sin(y_2)\cos(y_2)\cos(y_1)^2g_{10}^{y_2 y_2}(g_{10}^{y_4 y_3}- g_{10}^{y_5y_3})dx \wedge dy \nonumber\\&\wedge dy_1 \wedge dy_5 \wedge dy_4\big).
\end{align}
Here $g_{10}$ stands for the general metric ansatz chosen in (\ref{metricc}), (\ref{metric}), $L(u)$ is the $u$-dependent radius of the five sphere (\ref{metricc}), which is $1$ in the lowest order in $\alpha'$, $g_5$ is the metric of the internal AdS space and $g_{S_5}$ is the metric of the five sphere . The part of the  five form entering  
\begin{equation}
-\frac{\sqrt{-g} (F^{mag})^2}{2\times 5!}
\end{equation}
of the effective action for the metric components derived in Theorem \ref{theorem} is 
$
(F_5)^{mag}=*((F_5^{el})^0+(F_5^{el})^1).
$
The part of the $5$-form $F^+$, which enters the $\mathcal{T}$-tensor in (\ref{Ttensor}), is given by $F^+=(1+*)\big((F_5^{el})^0 +(F_5^{el})^1\big)$. Again we define
\begin{equation}
\mathcal{L}_{10}^{W}=\sqrt{\vert g_{10}\vert} \bigg[C^4+C^3 \mathcal{T}+C^2  \mathcal{T}^2 +C \mathcal{T}^3+\mathcal{T}^4\bigg].
\end{equation}
Since we consider a strong background field, it can therefore not be treated perturbatively. Each part of the higher derivative terms, which are schematically written above, will contribute to the EoM for the metric components. 
Knowing the solution for the metric in order $\mathcal{O}(\gamma^0)$, and now for $b=\frac{5}{4}$ on the interval $u \in [l,k]$, on which the Gauss-Lobatto grid (\ref{grid}) is defined, to high precision, allows us to compute \footnote{When we compute the variation, we are allowed to assume that the metric components abbreviated with $X \in \{L,U,\tilde{U},W,V\}$ do not depend on $x$, since terms of the form $\frac{\partial}{\partial \partial_x X}\mathcal{L}_{10}^{W}$, $\frac{\partial}{\partial \partial_x^2 X}\mathcal{L}_{10}^{W}$ must vanish, exactly as in the case $\mathcal{O}(\gamma^0)$. Otherwise the EoM for the gauge field $A_y= b x$ would get mass terms. In addition $A_y= b x$  is also a solution to the higher derivative corrected EoM for gauge fields.}
\begin{equation}
\frac{\partial}{\partial X(u)}\mathcal{L}^W_{10}, \hspace{0.8cm}\frac{\partial}{\partial X'(u)}\mathcal{L}^W_{10}\hspace{0.4 cm} \text{ and } \hspace{0.4 cm} \frac{\partial}{\partial X''(u)}\mathcal{L}^W_{10}
\label{terms}
\end{equation}
for $X \in\{U, \tilde U,W,V,L\}$
on said grid. This very tedious calculation can be abbreviated by the observation that the final result will only depend on $y_1$ and $y_2$ via the square root of the absolute value of the determinant of the metric. \\ \indent We define $\mathcal{L}_{10}$ to be (\ref{action2}) with $F_5^2$ being replaced by $2 \big((F_5)^{mag}\big)^2$. As before we  consider the system of  differential equations (\ref{diff}). The ansatz of $U$, $\tilde U$, $L$, $W$ and $V$ is the same as in (\ref{lasteq1})-(\ref{lasteq}) with the difference that $V \neq W$. The argument, why we could choose the higher derivative corrections to $W=V$ in the case $b=0$ to vanish, can now be only applied to either $W$ or $V$.  Without loss of generality we set
\begin{equation}
X(u)=\tilde{X}^0(u)e^{\gamma u^{d_X} \sum_{i=0}^M a^{X,M}_i c^M_i(x u-y)}
\end{equation}
for $X \in\{U, \tilde U, W,V ,L\}$ and 
 $a^{V,M}_i =0$ henceforth. We again write (\ref{diff}) as a $(4M+4) \times (4M+4)$ matrix equation $ A \cdot v = \chi$, where $A$, $v$ and $\chi$ are defined analogously as in section \ref{alt}. The requirement that the metric induced on the boundary is the Minkowski metric  gives $d_X>0$ for $X \in\{U, \tilde U,W\}$. With an analogous procedure as in section \ref{alt} we obtain that $d_L>1$. We determine the solution for several values of $l$, $M \in\{m_1,\dots,m_2\}$ and for different values for $k$ in the vicinity of $1$\footnote{Divergences of several terms in the non-simplified version of  (\ref{terms}), which cancel analytically, if we would expand them around the horizon, but not numerically due to finite machine precision, make it also impossible to choose $k=1$.} to ensure that the numerical error we commit, due to the fact that we cannot choose the interval $[l,k]$ arbitrarily close to $[0,1]$\footnote{This would require an explicit, analytic near boundary analysis of (\ref{terms}), which  is rather hopeless.}, doesn't cause unacceptably large errors in the following calculations (see section \ref{secgamma}). 
Requiring that there is no conical singularity at the horizon gives a correction factor to the temperature for a background-field-parameter $b= \frac{5}{4}$ of
\begin{equation}
 \bigg(1+\frac{\gamma }{2} \big(\frac{d}{d \gamma}(U(u)-\tilde{U}(u))|_{\gamma \to 0 , u\to 1} \big)  \bigg)\approx (1+\gamma 294.9).
 \label{tempcor}
\end{equation} In figure (\ref{errorT}) we computed the deviation from the average value of  the $\alpha'^3$-correction $T^\gamma$ to the temperature 
\begin{equation}
\Delta T^\gamma(l) = \frac{1}{m_2-m_1+1}\sum_{M=m_1}^{m_2} \frac{T^\gamma(M,l)-\bar T^\gamma}{\bar T^\gamma},
\end{equation}
where $\bar{T}^\gamma$ is the average over all considered configurations $M \in\{m_1,\dots,m_2\}$ and $l \in [1,1.05,\dots,1.4]$, $m_1$ was chosen to be $10$, $m_2$ was chosen to be $23$, $k$ was kept fixed at $k=0.99$.
 The maximal relative difference between two results for the coupling corrected temperature corresponding to the various choices for $M$ and $l$  is
 \begin{equation}
 \delta T^{\text{max}}:=\frac{T^\gamma_{\text{max}}-T^\gamma_{\text{min}}}{\bar{T}^\gamma}=0.00565,
 \end{equation}
where both the minimal and the maximal value for $T^\gamma$ are taken in  the case $l=0.14$, the maximal $l$-value of our analysis. Finally let us consider the function 
\begin{equation} \delta T(l):= \frac{T^\gamma_{\text{max}}(l)-T^\gamma_{\text{min}}(l)}{\bar{T}^\gamma}\label{deltaT},
\end{equation} where $T^\gamma_{\text{max/min}}(l)$ is the maximal/minimal value for $T^\gamma$ 
we obtained for a certain $l$. The results are displayed in figure (\ref{errorT}). 
\begin{figure}
\center
 \includegraphics[scale=0.85]{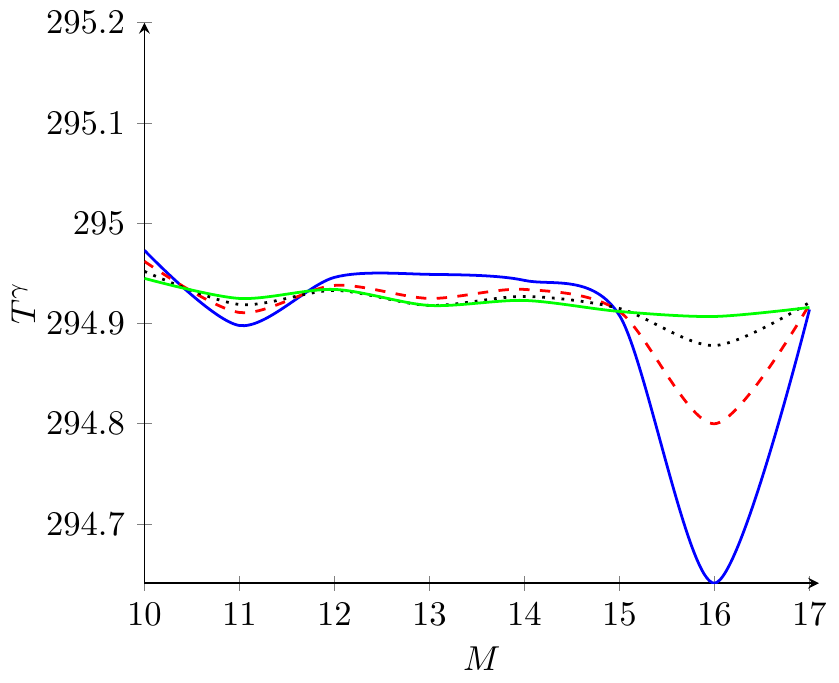}
 \caption{The higher derivative correction $T^\gamma$ for to the temperature, computed on  intervals $[0.1,k]$ for different values of $M$ (shown in a smoothed plot). The solid blue line shows the results for $k=0.975$, the dashed red line corresponds $k=0.98$, the dotted black line corresponds to $k=0.985$ and  the solid green line corresponds to $k=0.99$. The metric was extrapolated to $u=1$.}
 \label{errorTk}
 \end{figure} 
\FloatBarrier
In figure (\ref{errorTk}) we display the results for the correction factor to the temperature obtained by calculations on intervals $[0.1,k]$, we extrapolated the resulting coupling corrections to the metric to $u=1$.
 \subsection{Approximating higher derivative corrections to tensor QNMs without magnetic background field}
\begin{figure}
\includegraphics[scale=0.85]{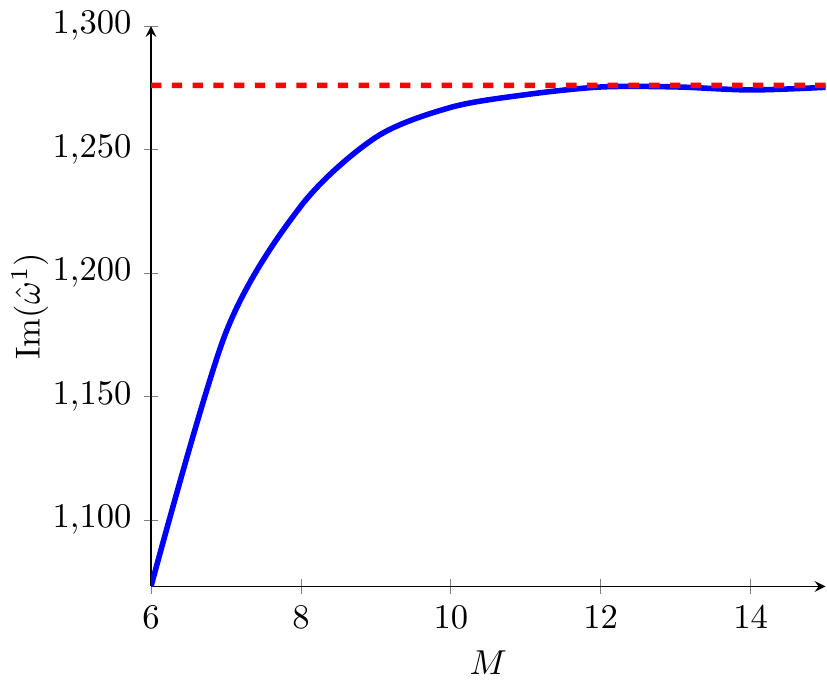}
\includegraphics[scale=0.85]{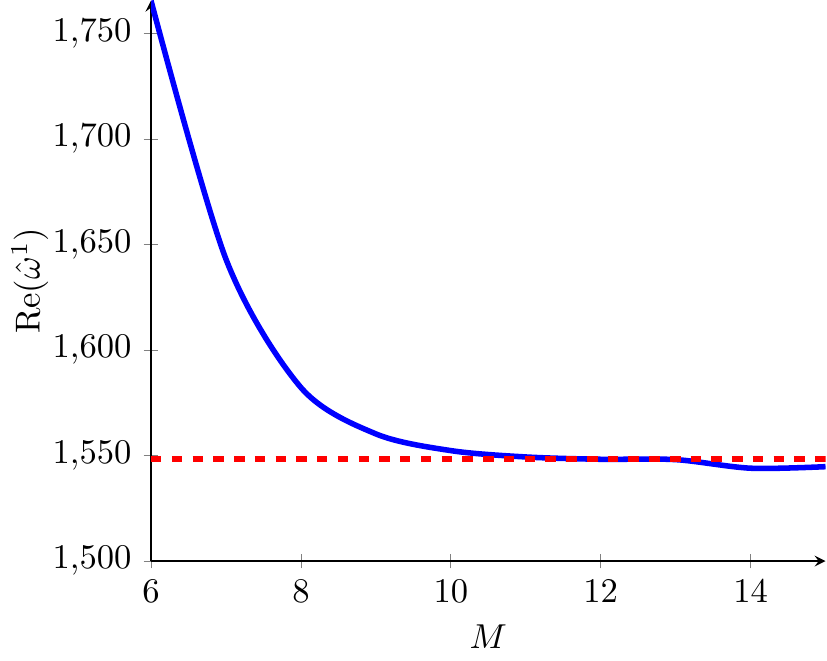}
\caption{The first order correction $\hat{\omega}^1$ to the lowest tensor-QNM frequency for $b=0$, $q=0$, computed via spectral methods on a grid $u \in [0.1,0.99]$ (solid blue line) compared with the exact result (red dashed line) for different values of $M$ shown in a smoothed plot.}
\label{QNMb=0}
\end{figure}

 Let us now  turn to fluctuations of the metric of a coupling corrected AdS-Schwartzschild black hole. Quasinormal modes  can be thought of as tiny perturbations of the geometry, which can be separated according to their transformation behaviour, respectively with the help of symmetry arguments. They are dual to quasiparticles on the field theory side and encode the response of the system to excitations around the equilibrium. We will consider tensor, or spin-$2$-fluctuations  $h_{x,y} $ in the $x,y$-plane with momentum in $z$ direction.
In this section we are going to approximate higher derivative corrections to these tensor QNMs  without considering background magnetic fields. The coupling corrections to spin-$2$-QNMs in this setup were first computed in \cite{S.Stricker}. Our aim is to reproduce these results by  applying a technique, which can be extended to derive coupling corrections to tensor QNMs of the coupling corrected magnetic black brane geometry, which we now know on an interval $[l,k]\subset [0,1]$. 
We consider the linearized differential equations obtained by varying the higher derivative corrected action with respect to fluctuations $h_{xy} dx dy$ of the background geometry. These EoM were first derived in \cite{Paulos2} and are given in the Appendix (\ref{EoMb=0}). The characteristic exponents of the differential equation (\ref{EoMb=0}) are given by $\pm \frac{i \hat \omega}{2}$, such that 
\begin{equation}
h=(1-u)^{-\frac{i \hat \omega}{2}} \phi(u)
\end{equation} 
where $\phi(u)$ is  regular  at the horizon and the exponent of $(1-u)^{-\frac{i \hat \omega}{2}} $ was chosen to correspond to infalling wave solutions. Here $\hat{\omega}$ is defined as $\hat{\omega}=\frac{\omega}{2 \pi T}$ to be consistent with the convention in \cite{S.Stricker}. In the case of $b=\frac{5}{4}$ we will use the convention $\hat{\omega}=\frac{\omega}{ \pi T}$, $\tilde{\omega}=\frac{\omega}{ r_h}$ to be consistent with \cite{K}. Considering the grid (\ref{grid}) again, we define the discrete differentiation matrix $A(M)$ as
\begin{equation}
A(M)_{ij}=\frac{2}{k-l}\partial_u  c_j\big|_{u \to u_i},
\end{equation}
where $c_j$ is the $j$-th Chebyshev cardinal function corresponding to the $j$-th grid point $u_j$. An alternative and numerically more convenient definition of $A(M)$ is given in  \cite{Boyd:Spectral}. Expanding $\phi$ in Chebyshev cardinal functions $\tilde{c}$ corresponding to the grid (\ref{grid}) in the form
\begin{equation}
\phi(u)=\sum_{i=0}^{M} \tilde c_i(u)a_i
\end{equation}
 allows us to formulate (\ref{EoMb=0p2}) as a matrix equation for the zero momentum mode $q=0$
\begin{equation}
O(M,\gamma,\hat{\omega})v=0
\label{MatrixEquation}
\end{equation}
with $v=(a_i)_{i \in \{0,\dots,M\}}$ and 
\begin{equation}
O(M,\gamma,\hat{\omega})_{ji}=\sum_{l=0}^{M}f_2(u_j)A(M)_{jl}A(M)_{li}+f_1(u_j,\gamma,\hat{\omega})A(M)_{ji}+f_0(u_j,\gamma,\hat{\omega})\delta_{ji},
\end{equation}
the function $f_i$ for $q=0$ are given in the Appendix \ref{Asecb=0}. We can split up $O(M,\gamma,\hat{\omega})$ as 
\begin{equation}
O(M,\gamma,\hat{\omega})=O^0(M,\gamma)+\hat{\omega}O^1(M,\gamma)+\hat{\omega}^2O^2(M,\gamma).
\end{equation}
This allows us to write (\ref{MatrixEquation}) as a generalized Eigenvalue problem
\begin{equation}
\begin{pmatrix}
O^0(M,\gamma)&O^1(M,\gamma)\\0&1
\end{pmatrix} \begin{pmatrix}
v\\ \hat{\omega}v
\end{pmatrix}=\hat{\omega}\begin{pmatrix}
0&-O^2(M,\gamma)\\1&0
\end{pmatrix} \begin{pmatrix}
v\\ \hat{\omega}v
\end{pmatrix}.
\label{gep}
\end{equation}
The idea is to solve (\ref{gep}) for $\hat{\omega}$ exactly in $\gamma$ with 
\begin{equation}
\gamma = \frac{\zeta(3)}{8}1000^{-3/2} \frac{1-\cos(\frac{\pi n}{\tilde M+1})}{2}.
\end{equation}
We chose $n \in \{0,\dots,\tilde{M}\}$, $\tilde{M}=80$. This Gauss-Lobatto grid corresponds to $\lambda$ values between $\lambda= \infty$ and $\lambda=1000$. The slopes at $\lambda=\infty$ of the curves of partially resummed coupling corrected results for $\hat{\omega}$ in the complex plane will give us the $\mathcal{O}(\gamma^1)$-corrections to the QNM  frequency $\hat{\omega}$.
 The boundaries of the interval on which the Gauss-Lobatto grid (\ref{grid}) lives are chosen to be $l=0.1 $ and $k=0.99$.
 We depict  the coupling corrections to the first QNM in figure (\ref{QNMb=0}). The results are displayed for different values of the grid size $M$ and show clear convergence towards the exact coupling corrections obtained in \cite{S.Stricker}. We plotted the first order coefficients of the $\gamma$-expansion of the first QNM frequency.
\begin{equation}
\hat \omega^1:= \partial_\gamma \hat\omega |_{\gamma=0}.
\label{corfac}
\end{equation}
Applying this method with $25 \ll M$ corrupts the results noticeably. For $10<M<20$ we obtain  good agreement with the already known results. Since the aim is to give a numerical approximation to the higher derivative corrections of tensor QNMs in the presence of a strong magnetic background field, that backreacts on the coupling corrected geometry, we take this result as motivation to apply this technique in the case $b=\frac{5}{4}$.
\FloatBarrier
 \subsection{Approximating higher derivative corrections to the first tensor QNM in the presence of a strong magnetic background field}
 \label{secgamma}
\begin{figure}
 \includegraphics[scale=0.8]{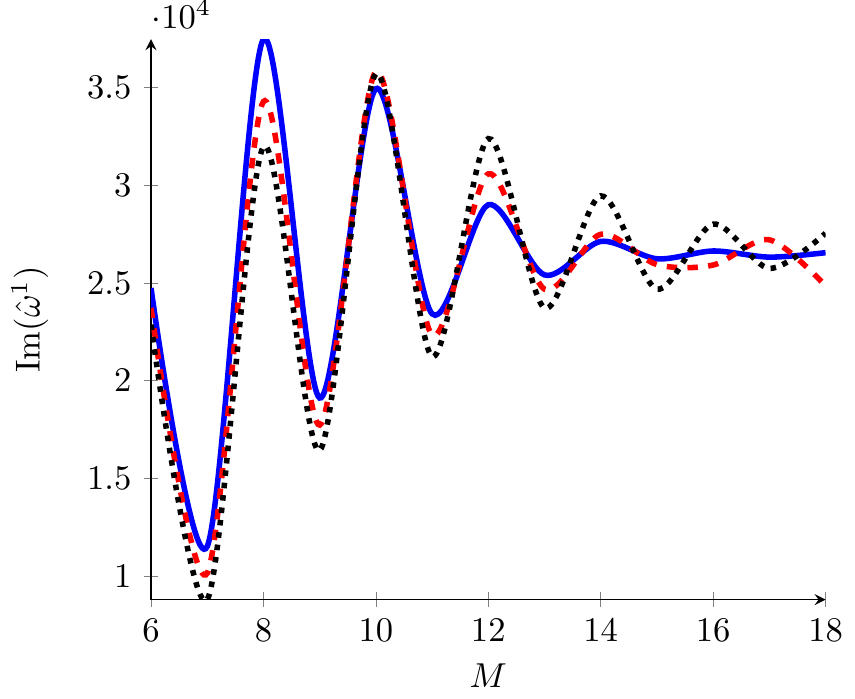}
  \includegraphics[scale=0.8]{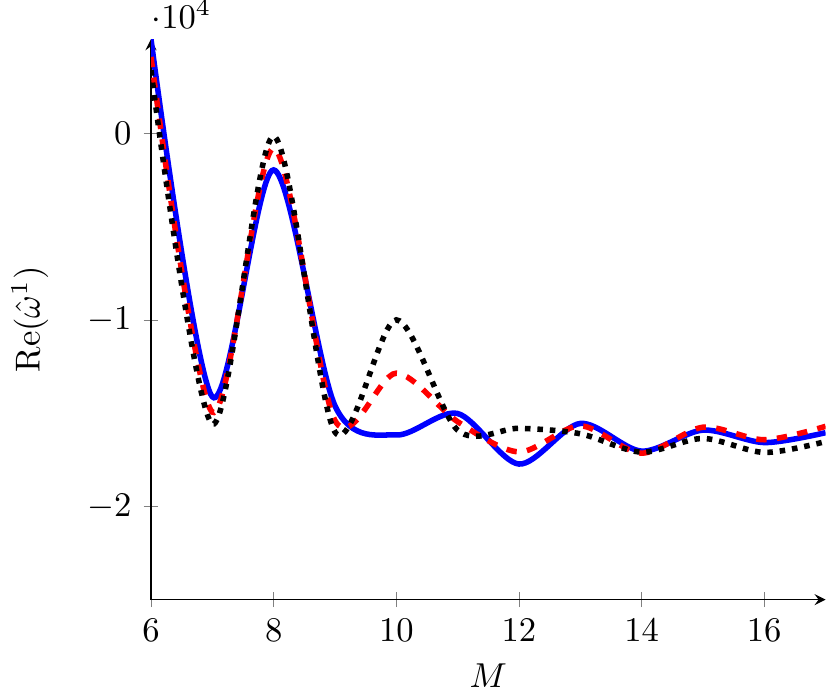}
  \caption{The convergence of the real and imaginary part of the correction  $\hat{\omega}^1$ for $b=\frac{5}{4}$ of the first tensor QNM computed for various grid sizes $M$ (shown in a smoothed plot) and for different interval sizes. The solid blue line corresponds to $[l,k]=[0.1,0.99]$, the dashed red line corresponds to $[l,k]=[0.11,0.98]$ and the dotted black line corresponds to $[l,k]=[0.12,0.97]$.}
   \label{compare}
 \end{figure}

\begin{figure}
 \includegraphics[scale=0.8]{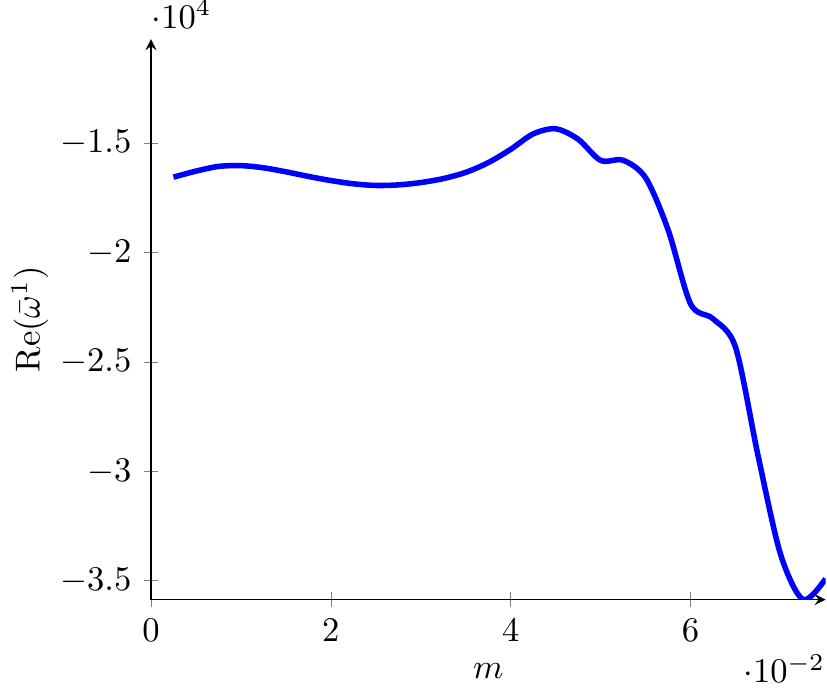}
  \includegraphics[scale=0.8]{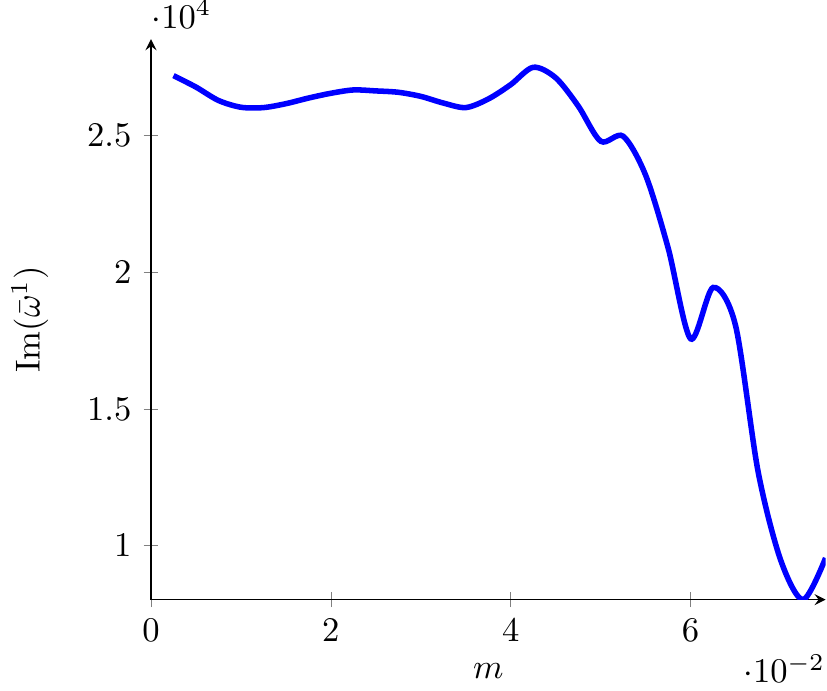}
  \caption{The convergence of  $\bar{\omega}^1(m)$ for different interval sizes $[0.1+m,0.99-m]$. The quantity $\bar{\omega}^1(m)$ is defined as the point of convergence of the $M$-dependent sequence $\hat \omega^1(M,m,m)$ with fixed $m$.}
   \label{compare2}
 \end{figure}

 The way we have chosen our background field together with considering  fluctuations  ensures that the linearized differential equations for $h_{x,y}$ decouple from those of other fluctuations.
 Our aim is to determine $\gamma$-corrections to the results  in 
 \cite{K}. As before the calculation is done for the case $q=0$.  We already have found the metric, respectively the functions $\tilde{U}, U, V, W, L$ up to order $\mathcal{O}(\gamma)$ for the parameter $b= \frac{5}{4}$ in the previous sections. The following metric ansatz describes tensor fluctuations of this geometry.
\begin{align}
ds_{\text{fluc}}^2= &- U(u)dt^2+ \tilde{U}(u) du^2+ e^{2V(u)}(dx^2+dy^2)+e^{2W(u)}dz^2+L(u)^2 \frac{4 b^2x^2}{3} dy^2\nonumber \\&+L(u)^2 \frac{2bx}{\sqrt{3}}dy \big(dy_3 \sin(y_1)^2 +dy_4 \cos(y_1)^2\sin(y_2)^2+dy_5 \cos(y_1)^2\cos(y_2)^2  \big)\nonumber \\ &
L(u)^2\big(dy_1^2+\cos(y_1)^2 dy_2^2+\sin(y_1)^2dy_3^2+\cos(y_1)^2\sin(y_2)^2 dy_4 +\cos(y_1)^2\nonumber \\ &\cos(y_2)^2 dy_5^2\big)+ h_{x,y}(u,t) dx dy.
\end{align}
Our strategy is very similar to the one of the previous chapters.  We choose  the same grids as before and evaluate the functions  
\begin{align}
 &\frac{\partial^2\mathcal{L}^{W,\text{fluc}}_{10}}{\partial^2 h_{x,y}}\bigg|_{h_{x,y} \to 0},\hspace{0.2cm} \frac{\partial^2 \mathcal{L}^{W,\text{fluc}}_{10}}{\partial^2( \partial_u h_{x,y})} \bigg|_{h_{x,y} \to 0},\hspace{0.2cm} \frac{\partial^2\mathcal{L}^{W,\text{fluc}}_{10}}{\partial( \partial_u h_{x,y})\partial h_{x,y}} \bigg|_{h_{x,y} \to 0}, \hspace{0.2 cm} \frac{\partial^2\mathcal{L}^{W,\text{fluc}}_{10}}{\partial^2 (\partial_{uu}h_{x,y})}\bigg|_{h_{x,y} \to 0} \nonumber \\ &\frac{\partial^2\mathcal{L}^{W,\text{fluc}}_{10}}{\partial (\partial_{uu}h_{x,y})\partial h_{x,y}}\bigg|_{h_{x,y} \to 0},\hspace{0.2cm} \frac{\partial^2\mathcal{L}^{W,\text{fluc}}_{10}}{\partial (\partial_{uu}h_{x,y})\partial(\partial_u h_{x,y})}\bigg|_{h_{x,y} \to 0},\hspace{0.2cm}\frac{\partial^2\mathcal{L}^{W,\text{fluc}}_{10}}{\partial^2(\partial_t h_{x,y})}\bigg|_{h_{x,y} \to 0},\hspace{0.2cm} \nonumber  \\&
 \frac{\partial^2\mathcal{L}^{W,\text{fluc}}_{10}}{\partial^2 (\partial_{tt}h_{x,y})}\bigg|_{h_{x,y} \to 0},\hspace{0.2cm}\frac{\partial^2\mathcal{L}^{W,\text{fluc}}_{10}}{\partial (\partial_{tt}h_{x,y})\partial h_{x,y }}\bigg|_{h_{x,y} \to 0},\hspace{0.2cm}\frac{\partial^2\mathcal{L}^{W,\text{fluc}}_{10}}{\partial (\partial_{tt}h_{x,y})\partial( \partial_u h_{x,y })}\bigg|_{h_{x,y} \to 0},\nonumber \\ & \frac{\partial^2\mathcal{L}^{W,\text{fluc}}_{10}}{\partial (\partial_{t}h_{x,y})\partial( \partial_{ut} h_{x,y })}\bigg|_{h_{x,y} \to 0}, \hspace{0.2 cm} \frac{\partial^2\mathcal{L}^{W,\text{fluc}}_{10}}{\partial^2 (\partial_{ut}h_{x,y})}\bigg|_{h_{x,y} \to 0}, \hspace{0.2 cm} \frac{\partial^2\mathcal{L}^{W,\text{fluc}}_{10}}{\partial (\partial_{tt}h_{x,y})\partial( \partial_{uu} h_{x,y })}\bigg|_{h_{x,y} \to 0}
 \label{func}
\end{align}
with \begin{equation}
\mathcal{L}^{W,\text{fluc}}_{10} =\mathcal{L}^W_{10}\bigg|_{ds_{10}^2 \to ds_\text{fluc}^2}
\end{equation} on the respective grid points. This very tedious calculation gives us the function given in (\ref{func}) as approximations in cardinal functions.
All other differentiations of $\mathcal{L}^{W,\text{fluc}}_{10}$ with respect to $h_{xy}$ vanish in order $\mathcal{O}(\gamma)$ for the linearized EoM. In the following we define 
\begin{equation}
\mathcal{J}(a,b):= \frac{\partial^2}{\partial(\partial_a h_{x,y} ) \partial(\partial_b h_{x,y} )}\mathcal{L}^{W,\text{fluc}}_{10}\bigg|_{h_{x,y}\to 0},
\end{equation}where we set $\mathcal{J}(0,b):=\frac{\partial^2 \mathcal{L}^{W,\text{fluc}}_{10}}{\partial(h_{x,y} ) \partial(\partial_b h_{x,y} )}\bigg|_{h_{x,y}\to 0}$.
Together with the Fourier transformed version of $h_{x,y}$
\begin{equation}
h_{x,y}(u,t) = \int \frac{d \omega}{2 \pi} \hat{h}(u,\omega) e^{ i \omega t}=:\int \frac{d \omega}{2 \pi} \hat{h} e^{ i \omega t}
\end{equation}
we can write 
\begin{align}
\int d^{10}x \mathcal{L}^{W,\text{fluc}}_{10}=& \text{vol}(S_5) \int \frac{d \omega}{2 \pi}\int du \int dx^3 \bigg(\frac{1}{2}\mathcal{J}(0,0) \hat {h}^2 +\mathcal{J}(u,0) \hat {h} \partial_u\hat {h}+ \frac{1}{2}\mathcal{J}(u,u) \nonumber \\ & (\partial_u\hat {h})^2 +\mathcal{J}(uu,u) \partial_{uu}\hat {h} \partial_u\hat {h}+ \frac{1}{2}\mathcal{J}(uu,uu) \partial_{uu}\hat {h} \partial_{uu}\hat {h}+\mathcal{J}(uu,0)\hat {h} \nonumber \\ &  \partial_{uu}\hat {h}+ \frac{\omega^2}{2} \mathcal{J}(t,t) \hat{h}^2+ \frac{\omega^4}{2} \mathcal{J}(tt,tt) \hat{h}^2-\omega^2 \mathcal{J}(tt,0) \hat{h}^2-\omega^2\mathcal{J}(tt,u)\nonumber\\ &  \partial_u \hat{h}+\omega^2 \mathcal{J}(t,ut) \hat{h} \partial_u \hat{h}+\frac{\omega^2}{2} \mathcal{J}(ut,ut)  (\partial_u \hat{h})^2 - \omega^2\mathcal{J}(tt,uu) \hat{h} \partial_{uu} \hat{h}\bigg)+\mathcal{O}(\hat{h}^3).
\end{align} 
A straightforward calculation shows that the rest of the  action can be written as\footnote{We omitted the prefactor $\frac{1}{2 \kappa}$ in front of the action, since it will not be important for the following calculations.}
\begin{equation}
\int d^{10}x \sqrt{-g_{\text{fluc}}}\bigg(R_{10}-\frac{8}{L(u)^5}-b^2\big(2 e^{-4 V(u)}+\frac{e^{-8 V(u)}}{2}h_{x,y}(u,t)^2 \big) \big(\frac{L(u)^2}{3}+\frac{2}{3 L(u)^6} \big)\bigg).
\end{equation}
We expand this action up to order $\mathcal{O}(\gamma)$ and up to order $\mathcal{O}(h_{xy}^2)$, which gives  terms of the form $\mathcal{L}^{\gamma=0}(h_{xy},\partial_{u}h_{xy}, \partial_{t}h_{xy}, \partial_{uu} h_{xy},\partial_{tt}h_{xy}, \partial_{ut} h_{xy}, u)$ as well as $\gamma \mathcal{L}^{\gamma}(\dots)$, with the same arguments. With the  Fourier representation of $h_{xy}$ we write the terms above in the same way as  $ \mathcal{L}_{10}^{W,\text{fluc}}$ depending only on $\hat{h}, \partial_u \hat{h}, \partial_{uu} \hat{h}, \omega ,u$.
With the variation of
\begin{equation}
\int d^{10}x \bigg(\mathcal{L}^{\gamma=0}+\gamma \mathcal{L}^{\gamma}+\gamma \mathcal{L}_{10}^{W\text{fluc}} \bigg)
\end{equation} with respect to $\hat h$  we end up with the EoM for $\hat{h}$ depending on the functions $\mathcal{J}(a,b)$ and the order $\mathcal{O}(\gamma)$ parts of the metric as an expansion in cardinals functions. Inserting the coupling corrected relation between the horizon radius and the temperature (\ref{tempcor}) shows that the characteristic exponents stay of the form $\pm \frac{i \hat{\omega}}{4}$. Here and in the following we will use the convention $\hat{\omega}=\frac{\omega}{ \pi T} $ and $\tilde{\omega}=\frac{\omega}{ r_h}$.\\\indent
Since we have to consider solutions that are infalling at the horizon we set again
 \begin{equation}
\hat{h}(u, \hat \omega)=(1-u)^{-\frac{i \hat{\omega}}{2}} \phi(u ,\hat{\omega}).
\end{equation}
We consider Gauss-Lobatto grids on intervals $[l,k]\subset [0.1,0.99]$ of size $M$ and approximate the function $\phi(u, \hat{\omega})$ by cardinal functions with expansion coefficients $a_{i}^M$. In analogy to the previous section the coupling corrected differential equation for $\phi$ in the presence of a strong magnetic background field is brought into the form (\ref{MatrixEquation}). The coupling correction to the  QNM is then again computed by considering this equation as a generalized Eigenvalue problem. We performed this calculation for various intervals $[l,k]$ and various grid sizes $M$\footnote{The higher derivative corrections to the metric were obtained by interpolations using cardinal functions on the interval $[0.1,0.99]$ and with $M=17$. We repeated the calculations displayed in figures (\ref{compare},\ref{compare2}) for metrics computed with various choices for $M$ and the interval (while we extrapolated to the full size of the interval on which we computed the QNM, if necessary) and found negligible differences regarding the final results.}.\\ \indent
 We define
$
\hat{\omega}^1(M,\tilde l,\tilde k)
$ to be the first order correction of the lowest tensor QNM computed with spectral methods on a grid with size $M$ living on the  interval $[0.1+ \tilde{l}, 0.99-\tilde{k}]$. The aim is to study how the results, towards which $\hat{\omega}^1(M,\tilde l,\tilde{k})$ converges for growing $M$, depend on the interval size. Figures (\ref{compare}) show the comparison between the $M$-dependent results for the real and imaginary part of $\hat{\omega}^1$ for different intervals $[l,k]$.
 The results for $\bar{\omega}^1(m)$, which is defined as the point of convergence with respect to $M$ of  $\hat{\omega}^1(M,\tilde l,\tilde{k})$ with $\tilde l = \tilde k =m$, are displayed in figure (\ref{compare2}). In general we observe  for relatively low values of $M$  reasonable convergence, similar to the case $b=0$, such that we can give an approximation of the higher derivative correction to the first tensor QNM with $b=\frac{5}{4}$.\footnote{The numerical errors of the following $\alpha'$-corrected QNMs were too large to give meaningful quantitative results. Regarding the size, as for the first QNM, their correction terms seemed to be one order of magnitude larger compared to the case of a vanishing background field.} We obtain $\hat \omega^1 \approx (-1.6+2.7 i)10^4$, such that 
\begin{equation}
\frac{\omega}{\pi T}\approx(2.0 - 4.7 i)+\gamma (-1.6+2.7 i)10^4+\mathcal{O}(\gamma^{4/3}).
\label{resQNM}
\end{equation}
The $\lambda \to \infty$ limit coincides with the findings in \cite{K}.
The correction $\gamma (-1.6+2.7 i)10^4$ to the lowest QNM in the case of very strong magnetic background field  is, similar to the higher derivative correction to the  temperature, one order of magnitude larger than in the case $b=0$. This is not surprising, but it  raises the question, whether it makes sense, to evaluate this coupling corrected first QNM at values for the 't Hooft coupling that would  correspond to a more realistic QCD limit $\lambda\approx 11$, which is obtained by naively choosing  
\begin{equation}
g_{\text{YM}}=\alpha_s|_{T \text{ large}} \approx 0.3
\end{equation} 
and $N=3$. Unlike in the case $b=0$ the sign of the real part of the first order correction  term is negative. In the next section we show that considering higher order corrections to the QNM coming from the first order correction to the EoM of $h_{xy}$ this behaviour is reversed already for small values of $\gamma$. For small values of $\lambda$ the real part of the first QNM for $b= \frac{5}{4}$ behaves similarly to the analogous quantity in the case $b=0$. 
\subsection{Resumming finite $\lambda$ corrections to the first tensor QNM in a strong magnetic background field}
\label{secgamma2}
\begin{figure}
\includegraphics[scale=0.9]{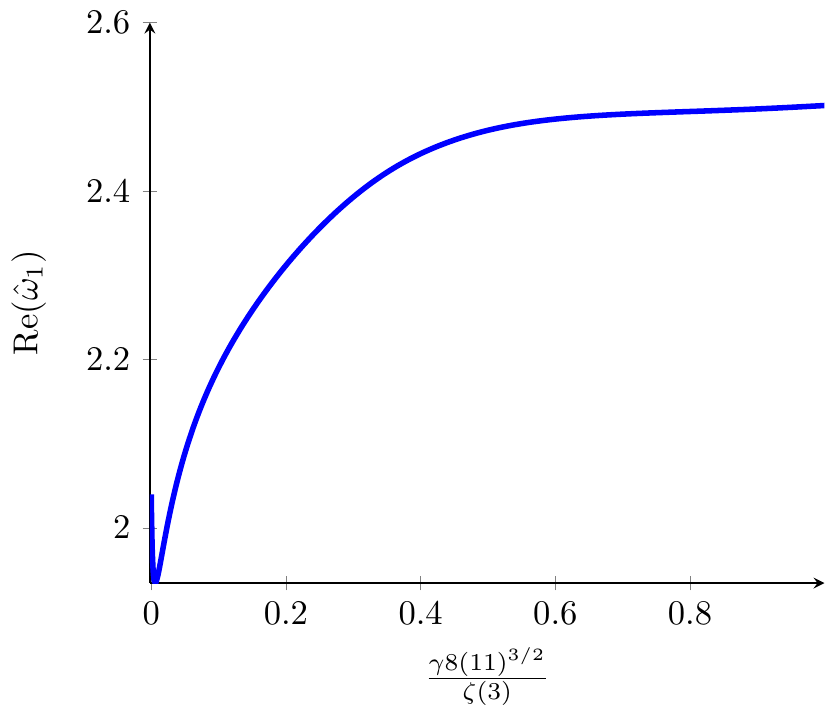}
\includegraphics[scale=0.9]{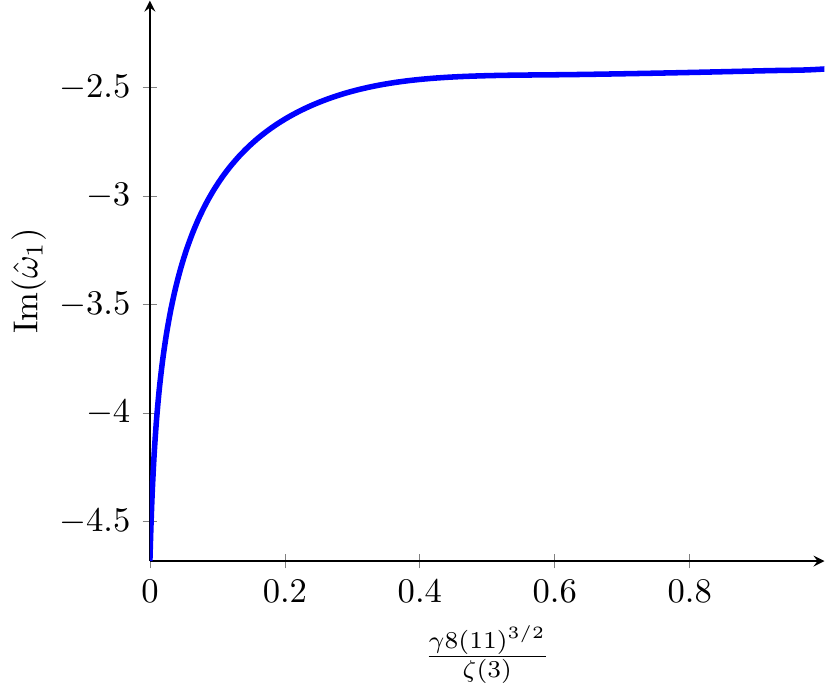}
\caption{The coupling corrected and resummed imaginary and real part of the lowest QNM with $b= \frac{5}{4}$ and $q=0$ averaged over different grid sizes and interval sizes. The maximal deviation from those suggest a negligible error for large $\lambda$, an error of $\approx 1\%$ for the imaginary part and $\approx 10 \%$ for the real part for $\lambda \to 11$. Interestingly the average values as well as the curves for large $M$ and large interval sizes $[l,k]$ converge for $\frac{\gamma 8 (11)^{3/2}}{\zeta(3)}\to 1$, or $\lambda \to 11$  to $2 b (1-i)$. The constant shape of the curve at small $\lambda$ suggests that this is also the  limit towards which the mode converges for $\lambda \ll 10$.
}
\label{fr1}
\end{figure}

\begin{figure}
\includegraphics[scale=0.9]{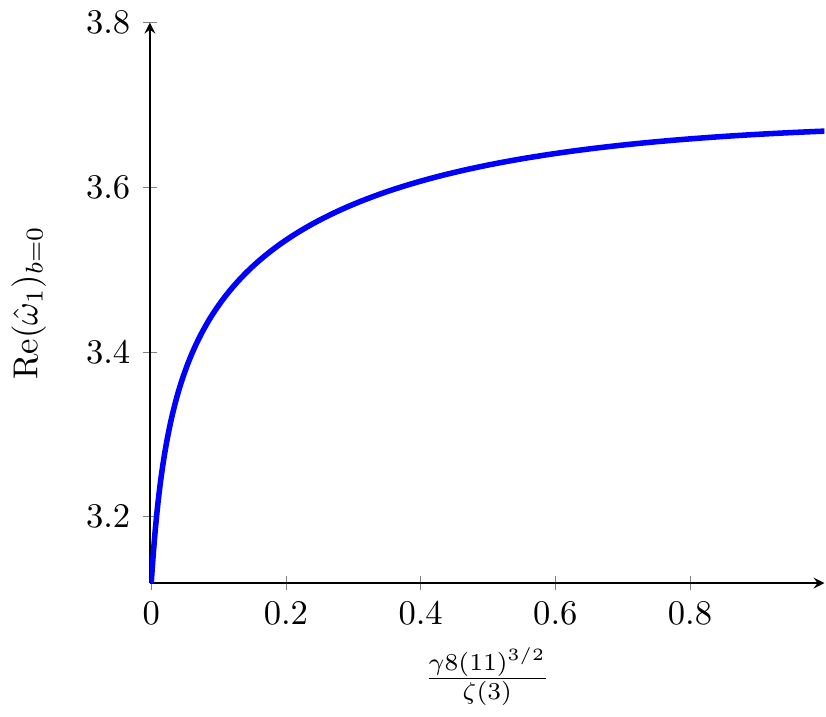}
\includegraphics[scale=0.9]{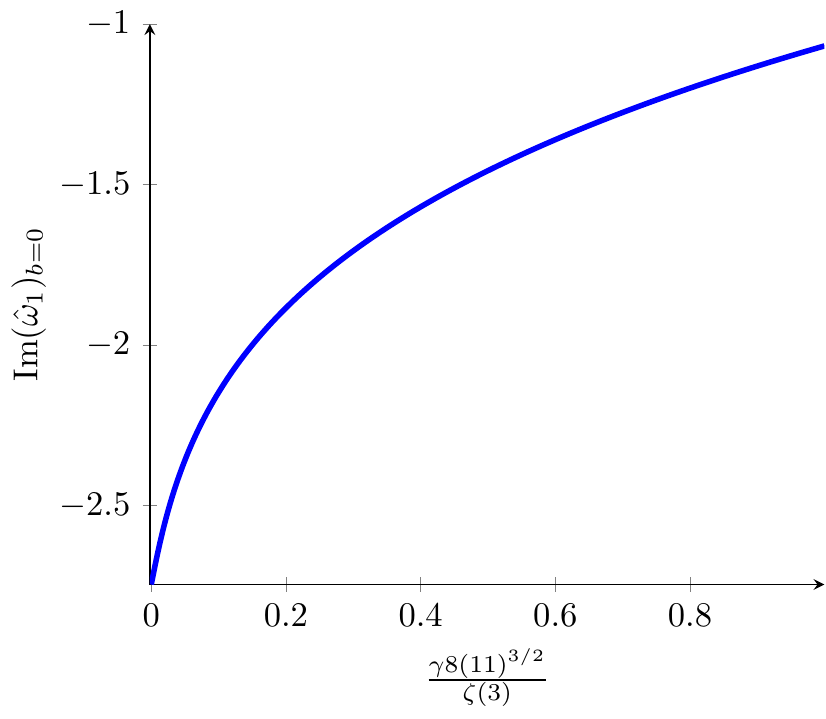}
\caption{The coupling corrected and resummed imaginary and real part of the lowest QNM with $b=0$ and $q=0$ on the same $\gamma$-interval as the plots shown in figure (\ref{fr1}). As in the $\lambda \to \infty$ limit \cite{K} the magnetic background field decreases both the real part of the QNM frequencies and the equilibration time $\tau \propto -\frac{1}{\text{Im}(\omega_1)}$. }
\label{fr2}
\end{figure}

\begin{figure}\centering
\includegraphics[scale=1]{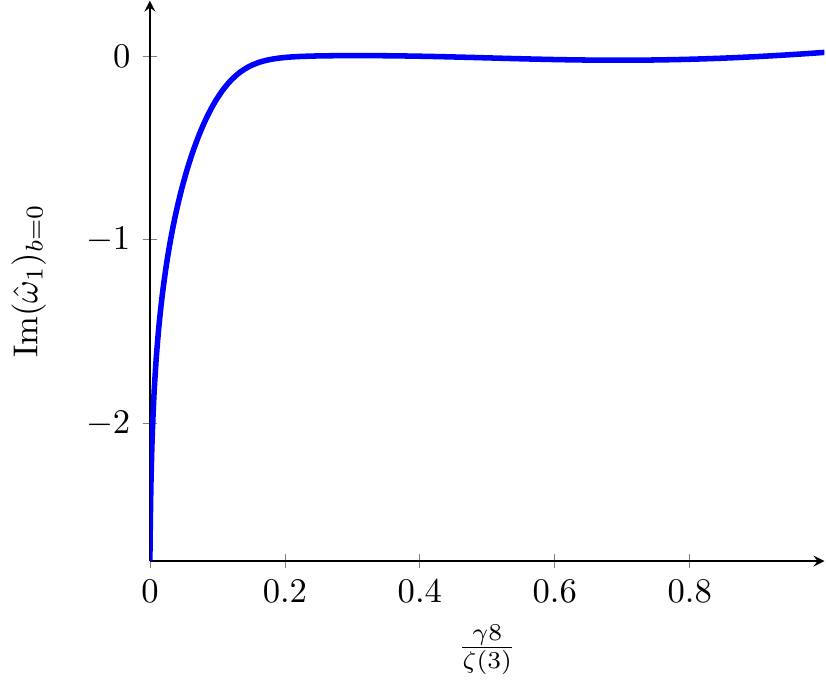}
\caption{The imaginary part of the coupling correction resummed first QNM frequency for $b=0$ on the $\gamma$-interval that corresponds to $\lambda \in[\infty,1]$. Unlike in the case of $b=\frac{5}{4}$ the imaginary part converges to $0$ here, reflecting that for vanishing interactions the equilibration time diverges. }
\label{fr3}
\end{figure}

\FloatBarrier
Finally we are going to consider  resummed coupling corrections to the first tensor QNM. Computing all higher derivative corrections of order $\alpha'^4$ to type IIb SUGRA or even higher orders dramatically exceeds current computational resources. However, there is a subset of higher derivative corrections in all orders $\alpha'^n$ to the QNM spectrum or to any other coupling corrected quantity computed within the AdS/CFT duality, that are already easily accessible, namely those that follow from the first order correction to the EoM of the corresponding field, in this case  $h_{xy}$.  Resumming these higher order corrections analogously to 
\cite{p2} will allow us to decrease $\lambda$ to almost arbitrarily small values without witnessing non-physical behaviour like an positive imaginary part of QNMs. Also the size of the resummed corrections is small compared to the $\lambda \to \infty$ spectrum for a wide range of $\lambda$ values. It should be added that this obviously  covers only one of many possible resummation schemes \cite{Buchel2} and that these partial resummations should be enjoyed with a grain of salt, as already pointed out in \cite{p2}. Their reliability at large $\gamma$ is uncertain and they should not be understood as exact predictions but rather as rough estimates, which, if taken seriously, should be tested with other equivalent  schemes. We postpone this additional analysis of this section to future work. Nonetheless the resummation we are going to present exhibits interesting features that we are going to discuss in the following.\\ \indent
We resum by truncating the EoM for $h_{xy}$ deduced in the previous section after the first order in $\gamma$ and compute exactly in $\gamma$ henceforth. In entire analogy to the calculation there, we apply spectral methods and write the task of finding the QNM spectrum as a generalized Eigenvalue problem, only that we are now interested in the resulting $\lambda$-curves of coupling correction resummed QNMs in the complex plane instead of their slope at $\lambda \to \infty$. We display the results in the figures (\ref{fr1}, \ref{fr2}, \ref{fr3}). The quantity $\hat \omega$ there is defined as $\hat \omega=\frac{\omega}{\pi T}$. We find that for small values of $\lambda$ both the imaginary and the real part of the first tensor QNM with $q=0$ and $b=\frac{5}{4}$ converge to a fixed value. In consistency with the $\lambda \to \infty$ results these values are smaller than  in the case $b=0$. For $b=0$ the imaginary part of the QNM converges to $0$ for small $\lambda$ (see figure (\ref{fr3})), whereas for $b=\frac{5}{4}$  it converges to $-2.5$ as seen in figure (\ref{fr1}). This is expected to happen, since without a background field and with very small 't Hooft  coupling nothing drives the equilibration of the QGP and the thermalization time, that can be estimated from the negative inverse of the imaginary part of the lowest QNM, diverges. The electromagnetic coupling doesn't approach zero  for small values of $\lambda$ (see e.g. (2.8) of 
\cite{Fuini} and corresponding footnote). Thus, in the case of a strong background field the QGP still equilibrates even if $\lambda$ is sent to small values\footnote{Any evaluation at $\lambda \to 0$ is neither feasible nor meaningful in this context. When we call $\lambda$ small, we mean $\lambda \ll 10$, $\lambda>0$.}, which is reflected by the comparison between the results displayed on the right hand side of figure (\ref{fr1}) and figure (\ref{fr2}). Out of caution it should be stressed that we treated only one possible channel. Therefore and because of the uncertain validity of partial resummations at small $\lambda$ our results suggest and don't prove this statement.
\FloatBarrier  
\section{Discussion}
In this work we provided a proof of the prescription found in \cite{p1}, regarding the higher derivative corrected five form in the presence of gauge fields,  for the special case of  a magnetic background field $F=b dx \wedge dy$. Using the higher derivative corrections to  the type IIb SUGRA action \cite{Paulos:2008tn} we computed the finite 't Hooft coupling corrected black brane metric, in which the strong background field back-reacts to the geometry. In this setting we found the 't Hooft coupling correction to the temperature (\ref{tempcor}) and computed the $\alpha'^3$ correction to the first tensor QNM (\ref{resQNM}). These correction terms turned out to be one order of magnitude larger than without a magnetic background field. The resummation of higher order corrections to this QNM frequency revealed an interesting pattern that reflects the intuitive expectation. For a vanishing background field and a vanishing 't Hooft coupling the imaginary part of the lowest (tensor) QNM frequency approaches $0$, this suggests that the equilibration time diverges in this case. For a strong background field of $b=\frac{5}{4}$ (which corresponds to $\mathcal{B}\approx 34.5 T^2$ for $\lambda \to \infty$) the imaginary part of the lowest QNM $\hat \omega(\lambda)$ converges to $-2.5$ for $\lambda \to 11$, which is the value for the 't Hooft coupling that naively corresponds to the QCD limit. The (coupling correction resummed) QNM frequency itself approaches $2b (1-i)$ for  $\lambda \to 11$. The form of the curve (\ref{fr1}) suggests that this is also the limit for $\lambda \ll 10$, indicating that the equilibration time of a QGP in a magnetic background field stays finite (and is of the same order of magnitude as in the $\lambda \to \infty$ limit) even if the 't Hooft coupling becomes extremely small.  \\ \indent It  should be added that there are many different ways to resum higher order corrections and that these resummations also should be taken with a grain of salt, when applied to compute quantities at large $\gamma$. They should be tested with other resummation schemes, otherwise the resummed results for small values of $\lambda$ have to be understood   as rough qualitative estimates at best.
\section*{Acknowledgements}
The author thanks Matthias Kaminski and Andreas Sch\"afer for useful discussions.  The author of this work was supported by the Research Scholarship Program of the Elite Network of Bavaria.
\section{Appendix}
\subsection{Equation of motion of tensor fluctuations for $b=0$}
\label{Asecb=0}
We define the function $h:=h_x^y$, such that one obtains \cite{Paulos2}
\begin{align}
\nonumber & h''(u) - \frac{u^2 + 1}{u (1 - u^2)} h'(u) + (\hat \omega^2 - \hat q^2 \frac{1-u^2}{u(1-u^2)^2} h(u) + 
\frac{\gamma}{4} \bigg( (3171 u^4 + 3840 \hat q^2 u^3 + 2306 u^2 \\ & - 600) u h'(u) + \frac{u}{(1-u^2)^2} \Big(600 \omega^2 - 300 \hat q^2 + \nonumber
         50 u + (3456 \hat q^2 - 2856 \hat \omega^2) u^2 + 
         768 u^3 \hat q^4 \\ & + (2136 \hat \omega^2 - 6560 \hat q^2) u^4 - (768 \hat q^4  + 
            275) u^5 + 3404 \hat \hat q^2 u^6 + 225 u^7\Big) h(u) +\nonumber\\ & 
    120 \frac{\hat \omega^2 - \hat q^2 (1 - u^2)}{u (1 - u^2)^2} h(u)\bigg)=0
    \label{EoMb=0}
\end{align}
from varying (\ref{actiongamma}) with respect to $h_{xy}$ in the case of a coupling corrected background metric with zero background fields. This differential equation simplifies to
\begin{equation}
f_2(u)\phi''(u)+f_1(u,\gamma,\hat{\omega}) \phi'(u)+f_0(u,\gamma,\hat{\omega}) \phi''(u)=0
      \label{EoMb=0p2}
\end{equation}
definig $h(u)=(1-u)^{-\frac{i \omega}{2}}\phi(u)$, where we set $q=0$. The coefficients $f_0$, $f_1$, $f_2$
 are given by \begin{align}
&f_0(u,\gamma,\hat{\omega})=((-2 \hat \omega (-2 i + 4 \hat{\omega} + u^2 \hat{\omega} + u (-2 i + 3 \hat{\omega})) + 
     \gamma (1 + u) (u^5 (-100 - 2306 i \hat{\omega})\nonumber \\ &\nonumber  + u^7 (450 - 3171 i \hat{\omega}) - 
        3171 i u^6 \hat{\omega} - 240 \hat{\omega}^2 + 100 i u^3 (i+ 6 \hat{\omega}) - 
        120 u^2 \hat{\omega} (-5 i + 12 \hat{\omega}) \\& + 2 u^4 \hat{\omega} (-1153 i + 2136 \hat{\omega}))) 
        \end{align}
        \begin{equation}        
 f_1(u,\gamma,\hat{\omega})= 2 (1 + u) (4 - 2906 \gamma u^4 - 865 \gamma u^6 + 3171 \gamma u^8 + 
        u^2 (4 + 600 \gamma - 4 i \hat{\omega}) - 4 i u \hat{\omega})
        \end{equation}

  \begin{equation}        
 f_2(u)= 8u(1+u)^2(-1+u).
        \end{equation}
\subsection{Expansion coefficients}
\label{expansion}
We give exemplarily the next order coefficients of the near horizon expansion of the magnetic black brane geometry without higher derivative corrections 
\begin{align}
u_3=&-\frac{1}{1080 l_0^{20} u_1}\bigg(-16 b^4 l_0^{24}-76 b^4 l_0^{16}-108 b^4 l_0^8-135 b^2 l_0^{22} u_1-540 b^2 l_0^{14}\nonumber \\& u_1+240 b^2 l_0^{20}+1440 b^2 l_0^{12}-720 b^2 l_0^4-675
   l_0^{20} u_1^2+4050 l_0^{18} u_1-3240 l_0^{10} u_1\nonumber \\&-5400 l_0^{16}+10800 l_0^8-6480\bigg)
\end{align}
\begin{align}
v_2=&-\frac{1}{360 l_0^{20} u_1^2}\bigg(11 b^4 l_0^{24}+11 b^4 l_0^{16}+8 b^4 l_0^8+45 b^2 l_0^{22} u_1+45 b^2 l_0^{14} u_1+30 b^2 l_0^{20}\nonumber \\&-180 b^2 l_0^{12}-270 l_0^{10} u_1+900
   l_0^8-720\bigg)
\end{align}
\begin{align}
w_3=&-\frac{1}{360 l_0^{20} u_1^2}\bigg(-7 b^4 l_0^{16}-8 b^4 l_0^8-45 b^2 l_0^{14} u_1+120 b^2 l_0^{12}-270 l_0^{10} u_1+900 l_0^8\nonumber \\ &-720\bigg)
\end{align}
\begin{align}
l_2=&-\frac{1}{1800 l_0^{19} \text{u1}^2}\bigg(-12 b^4 l_0^{24}+5 b^4 l_0^{16}+7 b^4 l_0^8-45 b^2 l_0^{22} u_1+45 b^2 l_0^{14} u_1\nonumber \\ &-60 b^2 l_0^{20}+120 b^2 l_0^{12}-60 b^2 l_0^4-1350 l_0^{18}
   u_1+1350 l_0^{10} u_1-2700 l_0^8+2700\bigg).
\end{align}

\end{document}